\documentclass[pra,twocolumn,amsmath,amssymb]{revtex4}

\usepackage[pdftex]{graphicx}
\DeclareGraphicsExtensions{.pdf, .jpg}
\usepackage{dcolumn}
\usepackage{bm}

\begin{document}
\author{M.~R. Tarbutt, J.~J. Hudson, B.~E. Sauer, and E.~A. Hinds}
\affiliation{Centre for Cold Matter, The Blackett Laboratory,
Imperial College London, London SW7 2AZ, UK}
\title{Preparation and manipulation of molecules for fundamental physics tests}
\begin{abstract}

\end{abstract}

\maketitle

\section{Introduction}
Atoms and atomic ions have long held a place at the very heart of
precision measurement and metrology. For example, the second is
defined in terms of the Cs hyperfine structure interval, the Rydberg
constant is measured by spectroscopy of atomic hydrogen and the
ratio of electron to proton mass is known from the oscillation
frequencies of trapped atomic ions. The importance of atoms in this
field lies partly in the detailed understanding that we have of
atomic structure but also in the technical capabilities that exist
for preparing and manipulating atoms and ions. In recent years, both
the computational methods for understanding molecules more fully and
the experimental methods for producing and controlling them have
advanced enormously. This has led to a surge of interest in using
molecules for precision measurements, especially where they offer
new properties that are not available from atoms and atomic ions.
For example, the rotational, vibrational and electronic structures
within a molecule offer a wider range of coexisting frequencies than
one finds in atomic systems. Moreover, polar diatomic molecules have
a built-in cylindrical symmetry, whilst more complex molecules can
have a handedness - structural conformations that atoms cannot
offer. In this chapter, we discuss some applications of molecules to
current problems in precision measurement and we outline recent
technical advances that make some of these applications possible.

\section{Testing invariance principles}

\subsection{Do fundamental constants vary in time?}
Recent measurements have shown that the expansion of the universe is
accelerating, requiring the Einstein equations of cosmology to have
a ``dark energy" term, which was previously assumed to be
zero\cite{Pee03a}. This surprise, together with the ongoing search
for a quantum theory of gravity, has led theorists to question some
of the most basic assumptions in our physical model of the universe,
including the assumption that the constants of nature are indeed
constant over time.

In atomic and molecular physics, two of these constants are
particularly important. They are the fine structure constant
$\alpha$, and the electron-to-proton mass ratio $\mu = m_e/m_p$. The
Rydberg energy $Ry$ sets the gross scale of electronic binding
energy. Relative to this, the fine structure splittings are
characteristically smaller by the factor $\alpha^2$ and the
hyperfine structure is smaller again by a further factor of order
$\mu$, since $\mu$ relates the magnetic moment of the nucleus to
that of the electron. Consequently, it is possible to search for a
variation of $\alpha$ by comparing fine and gross structure at two
different times. Similarly the variation of $\mu$ can be deduced
from a further comparison of the hyperfine structure with one of the
other energy scales. Molecules bring an important new dimension to
this search \cite{FlambaumAndKozlov07} by adding two more energy
scales: vibrational energy of order $Ry\sqrt{m_e/M}$ and rotational
energy of order $Ry(m_e/M)$, where $M$ is the reduced nuclear mass.
Uzan \cite{Uza03a} has recently reviewed both the theoretical
framework and the experimental tests for variable constants.

So far, the most productive experimental method has been to study
astronomical spectra, which permit measurements of $\Delta
\alpha/\alpha$ and $\Delta\mu/\mu$, typically with a precision of 1
part in $10^{5}$, over enormous time intervals of order 10~Gyr,
giving uncertainties of order $10^{-15}$/year in the average value
of $\dot{\alpha}/\alpha$ or $\dot{\mu}/\mu$. At this level, some
observations seem to hint at a variation\cite{Web01a,Rei06a}, while
others do not \cite{Sri04a,Kan05a}. In order to interpret the
astronomical data one needs to know the sensitivity of each
transition to $\alpha$ and $\mu$, which requires input both from
laboratory data and from numerical modelling. Using H$_2$ in this
way, Reinhold \emph{et al.} \cite{Rei06a} have found an average
variation over 12~Gyr of
$\dot{\mu}/\mu=(1.7\pm0.5)\times10^{-15}$/year, which differs from
zero by $3.4\sigma$. The ground state $\Lambda$-doublet of OH has
also been observed at large redshift. The transition frequency for
this interval relative to its hyperfine structure yields information
on both $\alpha$ and $\mu$ \cite{Kan05a}. When compared with the
structure of OH today, these astrophysical measurements looking back
6.5~Gyr indicate with $2\sigma$ confidence that
$\dot{\mu}/\mu<2.1\times10^{-15}$/year on average and is consistent
with zero.   A laboratory experiment using a molecular decelerator
to make slow OH \cite{Hud06a} (there is more on deceleration in
Sec.\,\ref{Sec:AG}) has recently improved our knowledge of these
transition frequencies in the current epoch. This will allow further
improvement in the astronomical data to yield an even better
determination of $\dot{\mu}$ and $\dot{\alpha}$ in the near future.
Another frequency with high sensitivity to $\mu$ is the inversion
splitting of NH$_3$ due to tunnelling.  Analysis of quasar
absorption spectra, comparing the inversion line to rotational
transitions, has placed a limit $\dot{\mu}/\mu = (-1\pm 3)\times
10^{-16} \mathrm{yr}^{-1}$ \cite{Fla07a} on the variation of $\mu$.
This leaves the astronomical measurements in an intriguing but
uncertain state. Are there systematic errors yet to be uncovered? Do
the constants vary or not, and if they do, is the variation
irregular in time or perhaps non-uniform in space?

There is a good prospect that laboratory measurements can help to
clarify the situation. Although the time intervals available for
comparison are short - years rather than giga-years - this method
provides a promising alternative because recently developed
frequency comb methods can link optical frequencies directly to the
Cs standard, giving an absolute frequency accuracy below one part in
$10^{15}$ \cite{FrequencyComb}. Even higher accuracy, approaching
one part in $10^{17}$ is becoming available in measuring relative
frequencies \cite{Dau05a}. Several molecular experiments of this
type are now under way.  It has been suggested \cite{Sch05a} that
the cold, trapped molecular ions H$_2^{+}$ or HD$^{+}$ could be used
to measure changes in $\mu$ to a precision of 1 part in $10^{15}$. A
second laboratory-based proposal \cite{Vel04a} is to compare the
inversion splitting measured in a slow, cold fountain of ND$_3$ to
an atomic reference. A third project is in progress to measure
vibrational transitions in SF$_6$ \cite{Amy05a}.

\subsection{Testing fundamental symmetries}\label{subsec:SymmetryTests}

The electromagnetic forces that bind atoms and molecules together
obey Maxwell's equations and the Dirac equation, as synthesised in
quantum electrodynamics. This field theory has three important
symmetry properties: it is invariant under space inversion (parity,
P) charge conjugation (exchange of particles and antiparticles, C)
and time reversal (T). These symmetries have profound experimental
consequences. For example, the eigenstates of atoms and molecules
have definite parity (unless there are degenerate conformations),
leading to selection rules for radiative transitions. For similar
reasons, atoms and molecules cannot have a permanent electric dipole
moment (EDM) (barring degenerate conformations). For example, the
well known EDM of ammonia is not permanent because the nitrogen atom
tunnels back and forth at a frequency set by the splitting of the
two opposite-parity field-free states. Of course, this splitting is
small, and therefore it takes only a modest electric field to induce
a dipole moment.

For some years it was surmised that all interactions possess these
symmetries, but an experiment in 1956 \cite{Wu56} showed that weak
interactions violate parity, as seen by the fact that radioactive
decay particles are emitted with a large left-right asymmetry.
Within a decade, an experiment on the decay of kaon particles showed
that strong interactions are also not symmetrical, having a small
asymmetry under the combined operation CP \cite{Christensen64}.
There is a theorem for the type of theories used to describe
particle interactions (local, Lorenz-invariant field theories) that
they must be invariant under the triple reflection CPT. Since CP
symmetry is broken, this theorem seems to indicate that T symmetry
is also broken at the same level. These symmetry (and asymmetry)
properties have played an important role in developing the standard
model of particle interactions, which describes electromagnetic,
weak, and strong interactions.

The discovery of CP-violation led to a fascinating question. Is it
possible that particles, atoms and molecules do have permanent
electric dipole moments after all? This would require interactions
that violate both P and T, but we know that the weak and strong
interactions together can do that. As it happens, the standard
model, with its standard P and T violation, predicts exceedingly
small EDM values. This is the result of a fortuitous cancellation,
which comes about from the simplicity of the standard model. The
pressing issue today is to discover what lies beyond the standard
model. In order to understand more clearly the origin of mass and in
order to accommodate a quantum theory of gravity it seems very
likely that there are more particles than the standard collection.
Such an increase in complexity immediately leads to the prediction
that particles, atoms and molecules have EDMs much larger than the
standard model values. They are still very small, but they are no
longer too small to measure. Therefore, the search for a permanent
electric dipole moment of an atom or molecule is really the search
for particle physics beyond the standard model. Molecules are
beginning to play a central part in this search.

The EDM $d_{e,\,p,\,n}\vec{\sigma}$ of an electron, proton or
neutron, is neccessarily aligned along the spin direction
$\vec{\sigma}$ of the particle. In essence, an EDM measurement in an
atom or molecule involves polarizing the system with an applied
external electric field and searching for the interaction $\eta d_x
\vec{\sigma} \cdot \vec{E}$ between the electronic or nuclear EDM
and the polarised atom/molecule. Schiff's theorem \cite{Sch63a}
states that $\eta=0$ if the atom/molecule is made of point particles
bound by electrostatic forces. In other words, the electronic or
nuclear EDM does not see the applied field because it is shielded
out by the other charged particles. This theorem is important for
its loopholes: nuclei are not point particles and the electric
dipole interaction is not screened when the electrons are
relativistic. Consequently, $\eta$ is not zero if the atom/molecule
is well chosen \cite{San65a,San75}. For example the best measurement
of the proton EDM comes from a measurement on TlF molecules
\cite{Cho89a}, where the large size of the Tl nucleus ends up giving
$\eta\sim 1$ for the nuclear spin EDM interaction. The upper limit
on the neutron EDM is known both directly, from measurements on free
neutrons \cite{Bak06a}, and indirectly, from nuclear spin
measurements on Hg atoms \cite{Rom01a}.

For electron EDM measurements, as opposed to neutron or proton
measurements, $\eta$ can be much larger than 1 if the electron moves
relativistically within the atom or molecule. For example, the Tl
atom is very sensitive to the electron EDM $d_e$, with $\eta= -585$.
Currently, the best limit on the electron EDM is derived from such
an experiment on Tl atoms \cite{Commins02}. The effective
interaction $\eta d_e \vec{\sigma} \cdot \vec{E}$ is linear in
$\vec{E}$ because the polarisation of the atom is proportional to
the applied electric field. This polarisation is small for atoms in
laboratory strength fields because it derives from the mixing of
higher electronic states induced by the field and these are
typically $10^{15}$~Hz higher in energy. By contrast, a heavy polar
molecule is polarised by mixing rotational states, which are
typically only 10~GHz away, giving five orders of magnitude more
polarisability and a correspondingly larger $\eta$. The polarisation
due to rotational mixing stops increasing once the molecule is
largely aligned with the field and then $\eta\vec{E}$ approaches a
saturated value. For the YbF molecule, this value is an enormous
26~GV/cm~\cite{Kozlov:1998}.

A group at Imperial College, including the authors of this article,
are currently in the process of measuring $d_e$ using the YbF
molecule~\cite{Hud02a}. In comparison with the Tl atom, this
molecule gives roughly 500 times more EDM interaction energy,
whereas the interaction with magnetic fields is essentially the
same. Since stray magnetic fields constitute the primary source of
systematic errors this is a significant advantage. However, Tl beams
have much better statistical noise because they are much more
intense than YbF beams. At present the gain in sensitivity is
roughly offset by the loss in signal, and the YbF experiment is
taking data at a level of precision similar to the Tl
experiment~\cite{Hud02a}. In the new era of high-precision molecular
beam measurements that we are discussing here, the need for
brighter, colder, slower sources is a recurring theme that we
address again in the next section.

Other molecular approaches to EDM measurement are also being
pursued. The group of DeMille is aiming to measure the electron EDM
in a metastable $\Omega$-doublet of PbO \cite{Dem01a} using vapour
in a cell. The group of Cornell is investigating the possibility of
using trapped molecular ions\cite{Stu04a}, among which HfF$^+$
appears to be a promising candidate. There is also a proposal to
make a dense sample of some radical such as YbF and to measure the
magnetisation induced by aligning the electron EDMs in an applied
electric field~\cite{Kozlov06}.

We turn now to parity violation without T violation. This is very
well understood in the standard model of particle physics as a
normal feature of weak interactions. It is also well established in
atomic physics through the measurements on Cs atoms~\cite{Ben99a}.
However, it remains a fascinating topic in the context of molecular
physics, partly because it has not been observed in molecules but
mainly because chirality plays such an important role in chemistry.
The weak interaction is predicted to alter the energy spectrum
between enantiomers of chiral molecules. Indeed, it is still a
subject of debate whether the parity violation of weak interactions
has played any role in establishing the chirality of the
biochemistry in living organisms \cite{Quack02,Sandars02}. The most
intensively studied species are the methyl halides, CH-XYZ, where X,
Y, Z are three different halogens \cite{Sch02a}. The largest effect
is predicted to be a 50.8 mHz shift in the C-F stretching mode
between left and right handed versions of CHFBrI. By contrast, the
best experimental results \cite{Zis02a} reach a precision of 50Hz.
It appears that cold trapped molecules will be necessary to measure
weak interactions in these systems. Much larger enantiomer shifts,
in the range of several Hz, have recently been predicted for some
rhenium and osmium complexes \cite{LargePNC} and these may be
observable using supersonic beams.

 There is also nuclear physics interest in parity
violation because it plays a role in nuclear structure. A
particularly interesting possibility is that weak interactions in
nuclei can induce an anapole moment, a P-odd multipole that produces
no external field and corresponds in lowest order to a toroidal flow
of current within the nucleus.  An experiment is underway at Yale
University to measure the anapole moment of the $^{137}$Ba nucleus
using BaF molecules~\cite{Dem07a}.

The last symmetry we mention here is Lorentz invariance, which has
been a central plank of 20th century physics and has so far shown no
indication of being violated in nature. Even so, it is possible that
new physics, associated with quantum gravity at the Planck energy
scale, could lead to very small violations of Lorentz invariance in
the laboratory \cite{Kostelecky97}. For example, there might be a
change in the energy of an atom depending on the orientation of its
spin relative to some preferred direction, such as its velocity in
the rest frame of the universe. Very sensitive experiments of this
sort have already been performed using a variety of atomic clocks
\cite{Kostelecky99}. More recently, it has been pointed out that
diatomic molecules provide a new way to investigate Lorentz
invariance by orienting the internuclear axis of the molecule
relative to the proposed preferred direction. The symmetry violation
could then be read out as a shift in the energy, bond length,
vibration frequency, or rotation frequency \cite{Mul04a}.
Sensitivities for H$_2$ and HD and their cations have been
calculated in \cite{Mul04a}. The authors conclude that an experiment
to measure the ground state rovibrational transitions can improve
limits on some elements of the Lorentz tensor $c_{\mu\nu}$
\cite{KosteleckySME, Kostelecky97} by an order of magnitude. It is very likely
that other more polar molecules could be convenient to use and that
this area will develop as techniques for preparing, cooling and
trapping molecules progress further.

In the rest of this chapter, we discuss the production and detection
of intense molecular beams, particularly of heavy polar radicals. We
also describe methods of using pulsed molecular beams to map fields
in the beamline and to detect interactions through the quantum
coherences. Finally, we describe advances in slowing heavy polar
molecules with a view to trapping them. These rather practical
issues are the key points to be addressed if molecules are to
fulfill the fantastic promise that we have just outlined for
elucidating exotic fundamental physics.

\section{Beams of cold polar radicals}

All molecular beams must begin with a source. A good source will
provide a large number of molecules in the quantum state of interest
for the experiment. In most cases, the molecules need to be prepared
in a single low-lying rotational state, and so the temperature of
the molecules should ideally be smaller than the rotational energy
spacing, typically 1\,K or less. Pulsed sources of cold molecules,
very narrowly distributed in both position and velocity, offer many
advantages. They can be prepared with very high intensity without
imposing an excessive gas load on the vacuum system, they can be
used to map the electromagnetic field along the beamline with high
resolution and precision, they allow quantum coherences to be
prepared and manipulated with great accuracy, and they can be slowed
down in a Stark decelerator to increase the coherence times. In this
section, we concentrate on the formation and detection of cold,
pulsed beams of the radical molecules that are typically required
for measuring the P and T violating interactions discussed above.

Supersonic expansion is a very common technique for producing cold
molecular beams \cite{Scoles, Campargue}. A high pressure gas
expands through a nozzle into a vacuum chamber, acquiring a high
centre-of-mass velocity but very narrow velocity distribution. The
translational degrees of freedom are cooled, as are the internal
degrees of freedom of the molecules. While the first beams were
continuous, the method was later extended by using pulsed valves
with short opening times \cite{Gentry(1)78, GentryChapterScoles}.
The individual pulses could then be made very intense, without the
gas load becoming excessive. Often, the molecules of interest have
low vapour pressure, and need to be formed by laser ablation or
electric discharge techniques. This is usually done immediately
outside the nozzle of the pulsed valve, or inside an extended
nozzle, where the density of carrier gas is high enough to entrain a
useful fraction of the molecules produced \cite{Powers(1)97}. In our
laboratory, we have used laser ablation to produce cold beams of YbF
\cite{Tarbutt(1)02}, CaF and LiH molecules \cite{Tokunaga(1)07}. In
all cases, we detect our pulsed beams using time-resolved,
Doppler-free, laser induced fluorescence (LIF). This detection
method is very well suited for precision measurements where high
sensitivity, high frequency resolution, and good beam diagnostics
are all required.

\subsection{Apparatus}

\begin{figure}
\includegraphics{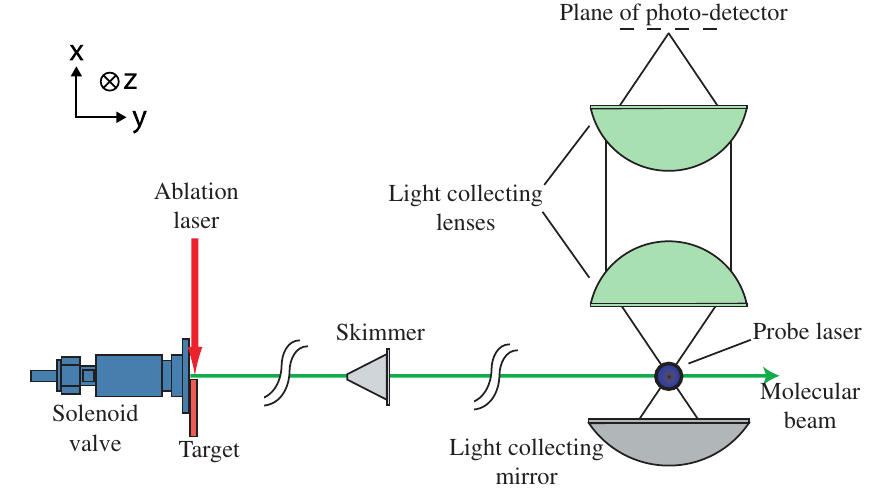}
\caption{Typical experimental setup for producing pulsed beams of
cold polar radicals, and detecting them by laser induced
fluorescence. Not to scale. \label{Fig:SupersonicSetup}}
\end{figure}

The typical experimental setup is illustrated in Fig.
\ref{Fig:SupersonicSetup}. A solenoid valve emits short pulses of
the carrier gas \cite{PulsedValve}, which is usually Ar, Kr or Xe,
into a vacuum chamber maintained at pressures below $10^{-4}$\,mbar.
The central part of the gas pulse passes through a skimmer, of
diameter 1-2\,mm, placed 50-100\,mm downstream of the source, and
into the high vacuum region where the pressure is below
$10^{-7}$\,mbar. A fast ionization gauge placed on the axis of the
beamline can be used to ensure good alignment. Using two such
gauges, one immediately outside the nozzle and the other much
further downstream, the initial width, the speed and the
translational temperature of the gas pulses can all be measured. The
shortest gas pulses we obtain with this valve have a full width at
half maximum (FWHM) of 81\,$\mu$s \cite{Tarbutt(1)02}. In an ideal
supersonic expansion from a reservoir at temperature $T_{0}$, the
carrier gas approaches a terminal velocity

\begin{equation}
v_{T} = \sqrt{(2k_{B}T_{0}/m)\gamma/(\gamma - 1)},
\end{equation}
where $\gamma$ is the specific heat ratio, 5/3 for an ideal
monatomic gas, and $m$ is the mass of an atom of the carrier gas.
Our measurements using a range of carrier gases and temperatures
show that the true speed of the carrier gas is 5-15\% faster than
$v_{T}$.

A suitable target, placed outside the nozzle of the valve, is
ablated using light from a Q-switched laser. To ensure minimal
disruption of the gas pulse, the target is thin in the direction of
the gas-jet, typically 2-5\,mm. The resulting ablation plume can
usually be observed by eye. It is forward peaked along the normal to
the target, which is usually perpendicular to the direction of the
beamline. Some of the atoms and molecules in the plume become
entrained in the high density carrier gas, thereby cooling to
temperatures and speeds that approach those of the carrier gas. The
target may either contain all the precursors needed to form the
molecules of interest, or some of the precursors can be added to the
carrier gas. For example, we have produced beams of YbF either by
ablating pure Yb and mixing SF$_6$ into the carrier gas, or by
ablating a pressed target containing a mixture of Yb and AlF$_{3}$
powders. The former method is slightly simpler to realise, but both
methods work equally well. We typically use ablation pulses of
5-10\,ns duration, focussed to a spot-size of 1-2\,mm. Under these
conditions, the optimal ablation energy is in the range 10-50\,mJ.

A photomultiplier, placed some distance (typically 10-150\,cm)
downstream, detects the molecules by means of cw-laser-induced
fluorescence. The laser is directed at right angles to the molecular
beam in order to minimise the Doppler shift and is tuned to a strong
molecular transition. The transit time of the molecules through the
laser beam is typically $5-10\,\mu$s, giving this method high
temporal resolution. The width of the fluorescence excitation
spectrum is usually greater than the transit time limit, being
typically 20-50\,MHz. This comes from a combination of the residual
Doppler width, due to the angular distribution of the molecules, and
the natural lifetime of the excited state. Such high spectral
resolution is typically needed in order to resolve the hyperfine
structure and hence to permit coherent state manipulation and
readout at rf frequencies. The fluorescence detection method is very
sensitive. If the detection efficiency is $\epsilon$ (assumed $\ll
1$) and there are $N$ molecules per shot passing through the
detector area in a time interval $w'$, then the shot noise limited
signal-to-noise ratio for a single shot will be $s:n = \epsilon
N/\sqrt{b w' + \epsilon N}$, where $b$ is the background rate of
detected photons. In a good detection setup, $b w' \approx 1$ and
$\epsilon$ is in the range 1-10\%. Taking $\epsilon = 0.02$ as a
typical value, we find that the signal-to-noise ratio per shot is 1
when $N = 81$ molecules.

\subsection{Translational temperature and source size}

\begin{figure}
\includegraphics{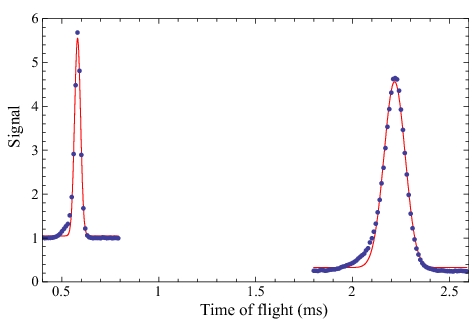}
\caption{Time of flight profiles of ground state YbF recorded by
two laser induced fluorescence detectors, situated 340\,mm and
1300\,mm from the source. The lines are Gaussian fits to the two
datasets. \label{Fig:Tofs}}
\end{figure}

Figure \ref{Fig:Tofs} shows the time-of-flight profiles of ground
state YbF molecules recorded by two separate LIF detectors placed
340\,mm and 1300\,mm from the pulsed source. The background laser
scatter is lower at the downstream detector. In this experiment, the
molecules were produced by ablating pure Yb just outside the nozzle
of the solenoid valve which was pressurized to 4\,bar with a mixture
of Ar(98\%) and SF$_{6}$(2\%). The probe laser was tuned to excite
the $F=1$ hyperfine component of the
$X^{2}\Sigma^{+}(v''=0)-A^{2}\Pi_{1/2}(v'=0)\,Q(0)$ transition.
There are four contributions to the spread of arrival times - (i)
the temporal spread of the molecules produced at the source, (ii)
their spatial spread at the source, (iii) the forward velocity
distribution in the pulse and (iv) the temporal resolution of the
detector. The last of these is usually small enough to be neglected.

The flux of molecules with speeds in the interval from $v$ to $v +
dv$ is usually taken to be $f(v)\,dv = A
v^{3}\exp(-M(v-v_{0})^{2}/2k_{B}T)\,dv$, where $M$ is the mass, $T$
is the translational temperature, $v_{0}$ is the central velocity
and $A$ is a normalizing constant. Consider those molecules born in
the source at time $t_s$ with initial position $s$ along the beam
axis. In the LIF detector, placed a distance $L$ away from the
source, these molecules produce a time-dependent signal

\begin{eqnarray}
h(t,t_{s},s)=&\frac{A(L-s)^{4}}{(t-t_s)^{5}}\exp\left(\frac{-M
v_{0}^{2}}{2k_{B}T}\frac{(t_{0}-s/v_{0}-t+t_s)^{2}}{(t-t_s)^{2}}\right),\label{detectorSig1}
\end{eqnarray}

\noindent where $t_{0}=L/v_{0}$. As we shall soon see, both the
temporal width and spatial width of the source are small, $t_s \ll
t_{0}$ and $s \ll L$. Furthermore, at typical detection distances,
the range of arrival times is much smaller than the mean arrival
time, and it is valid to set $t \approx t_{0}$ everywhere except in
the numerator of the exponent. With these approximations,
Eq.\,(\ref{detectorSig1}) simplifies to

\begin{equation}
h(t,t_s,\rho) \approx A \frac{L^{4}}{t_{0}^{5}}
\exp\left(-\frac{4\ln2(t-t_{0}-t_s-\rho)^{2}}{w^{2}}\right),
\end{equation}

\noindent where $\rho = - s / v_{0}$, and $w = \left(8\ln2\,k_{B}T
t_{0}^{2}/\left(M v_{0}^{2}\right)\right)^{1/2}$ is the temporal
width (FWHM) of the pulse due to the thermal spread of forward
velocities.

The signal at the detector is obtained by integrating over the
temporal and spatial distributions present in the source. We do not
know these distributions, but we can hope to obtain a measure of
their characteristic widths. In this spirit, we assume that the
source emits the normalised distribution

\begin{equation}
g(t_s,\rho) = \frac{4\ln2}{\pi \Delta_{t_s}
\Delta_{\rho}}\exp\left(\frac{-4\ln2\,t_s^{2}}{\Delta_{t_s}^{2}}\right)\exp\left(\frac{-4\ln2\,\rho^{2}}{\Delta_{\rho}^{2}}\right).
\end{equation}

\noindent Then, the detector records the signal

\begin{align}
h(t) &= \iint h(t,t_s,\rho)\,g(t_s,\rho)\,dt_s\,d\rho \nonumber \\
&= A \frac{L^{4}}{t_{0}^{5}} \frac{w}{w'}
\exp\left(-4\ln2\frac{(t-t_{0})^{2}}{w'^{2}}\right)
\end{align}

\noindent where the pulse width
$w'^{2}=w^{2}+\Delta_{t_s}^{2}+\Delta_{\rho}^{2}$ includes the
broadening due to the distribution of positions and times where
molecules are first formed.

As shown by the Gaussian fits in Fig.\,\ref{Fig:Tofs}, the recorded
profiles are well described by the model except in the high velocity
tails. The long tail, indicating a hotter component in the beam, is
a common feature of our source when optimized for maximum signal and
minimum shot-to-shot fluctuation; it can be removed by re-optimizing
for low temperature. By using two well-separated detectors, the
translational temperature can be obtained. It is

\begin{equation}
T = \frac{M
v_{0}^{2}}{8\ln2\,k_{B}}\frac{w_{2}'^{2}-w_{1}'^{2}}{t_{2}^{2}-t_{1}^{2}},
\end{equation}

\noindent where $w_{1}'$, $w_{2}'$, $t_{1}$ and $t_{2}$ are the
widths (FWHM) and central arrival times obtained from Gaussian fits
to the downstream and upstream data. For the data shown, the speed
is $v_{0}=586$\,m/s and the temperature is $T=4.8$\,K. The use of
two detectors allows an unambiguous determination of the
temperature, whereas a single detector alone sets only an upper
limit. However, if the detector is far from the source, that upper
limit can be very close to the true temperature. For this data, the
downstream detector alone provides an upper limit that is just 1.3\%
higher than the measured $4.8$\,K.

Having measured the temperature, we can also extract from the data
upper limits on the initial temporal and spatial spreads at the
source. For the data shown, these are $\Delta_{t_s} < 14.5\,\mu$s
and $v_{0}\Delta_{\rho} < 8.5$\,mm. In separate experiments, using a
dual ablation technique, we have measured $\Delta_{t_s} \approx
5\,\mu$s \cite{Tokunaga(1)07}.

\subsection{Molecular flux}\label{subsec:molecular flux}

We consider next how to determine the absolute flux of molecules
from the LIF signal. The detector counts the flux of photons, and we
wish to convert this to a flux of molecules by knowing the mean
number of fluorescent photons emitted from each molecule. To find
this, we model the molecule as a 3-level system and use rate
equations. Every molecule starts out in level 1 and passes through
the laser, which excites the resonance between levels 1 and 2. The
excitation rate is $R$, as is the rate of stimulated emission. It is
directly proportional to the laser intensity, $I$, and depends on
the detuning $\delta = \omega_{L} - \omega_{12}$ of the laser
angular frequency, $\omega_{L}$, from the molecular resonance
frequency, $\omega_{12}$. Taking the optical Bloch equations in the
limit where the coherences have reached a steady state, one finds
that
\begin{equation}
R = \frac{\Gamma/2}{(1+4\delta^{2}/\Gamma^{2})}\frac{I}{I_{s}},
\label{Eq:ExcitationRate}
\end{equation}

\noindent where $\Gamma$ is the spontaneous decay rate of level 2,
$I_{s} = \epsilon_{0} c \hbar^{2} (\Gamma/2)^{2}/D^{2}$ is the
saturation intensity, and $D$ is the matrix element of the dipole
operator connecting levels 1 and 2. Level 1 is stable, while level 2
decays with rate $r \Gamma$ to level 1, and rate $(1-r)\Gamma$ to
level 3 which represents all the other states in the molecule. The
rates for excitation or decay out of level 3 are negligible.

We solve the rate equations to find the number of molecules in
level 2 as a function of time, $N_{2}(t)$. Integrating $\Gamma
N_{2}(t)$ over the laser-molecule interaction time, $\tau$, we
find the number of fluorescent photons emitted per molecule to be

\begin{equation}
n_{p} =
\frac{R\,\Gamma}{R_{+}-R_{-}}\left(\frac{e^{-R_{+}\tau}-1}{R_{+}}
-
\frac{e^{-R_{-}\tau}-1}{R_{-}}\right),
\label{Eq:photonsPerMolecule}
\end{equation}

\noindent where

\begin{equation}
R_{\pm} = R + \Gamma/2 \pm \sqrt{R^{2}+ r\,R\,\Gamma +
\Gamma^{2}/4}. \label{Eq:Rpm}
\end{equation}

In the limit where $R_{+}\tau \gg 1$ and $R_{-}\tau \gg 1$,
$n_{p}$ acquires its asymptotic value $n_{p,{\rm max}} = 1/(1-r)$;
this is simply the sum of the obvious geometric series, $n_{p,{\rm
max}} = \sum^{\infty}_{N=0} r^{N}$. For good detection efficiency,
we would like to ensure that $n_{p}$ reaches its maximum possible
value, and so we should examine how easily this limit is reached.
If, as is usual, level 2 is an electronically excited state with
an allowed electric dipole transition, $\Gamma$ will typically
exceed $10^{7}\,s^{-1}$. The interaction time, $\tau$, is usually
greater than 1\,$\mu$s, and so we are in the limit $\Gamma \tau
\gg 1$. If we suppose that the laser excitation is weak, in the
sense that $R \ll \Gamma$ (or, equivalently, $I \ll I_{s}$), we
get

\begin{equation}
n_{p} = \frac{1-e^{-R(1-r)\tau}}{1-r} \quad (\Gamma\tau \gg 1, R
\ll \Gamma).
\end{equation}

\noindent The limit $n_{p}=1/(1-r)$ is reached once $R(1-r)\tau \gg
1$, at which point the fluorescence signal saturates. Note that it
is usual for $\Gamma \tau$ to be greater than 100, and that for
molecules, it is rare to have $r$ close to 1. It follows that the
above `saturation' condition can easily be met, even when $I \ll
I_{s}$ -- the fluorescence signal saturates for laser intensities
well below $I_{s}$ because the interaction time is very long
compared to the time required for the molecule to reach level 3, the
dark state.

\begin{figure}
\includegraphics{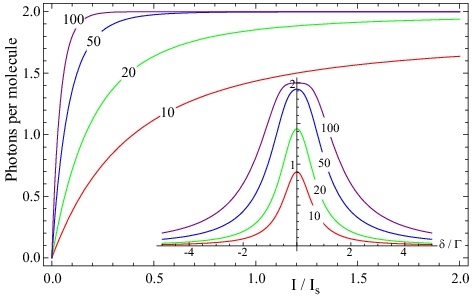}
\caption{Number of fluorescent photons per molecule, $n_{p}$, when
the branching ratio for returning to the initial state is $r=0.5$.
In the main graph, $n_{p}$ is plotted versus the intensity,
$I/I_{s}$, for $\Gamma\,\tau = 10,20,50,100$. The inset shows for
each case how $n_{p}$ varies with the laser detuning,
$\delta/\Gamma$, when $I=I_{s}/3$. \label{Fig:Saturation}}
\end{figure}

Figure \ref{Fig:Saturation} shows the value of $n_{p}$ as a function
of $I/I_{s}$ for four different values of $\Gamma \tau$ with
$r=0.5$. The maximum value of $n_{p}$ is 2, and in the case of
$\Gamma \tau = 100$, $n_{p}$ is within 1\% of this limit at
$I=I_{s}/4$. At lower values of $\Gamma \tau$, the asymptotic value
is lowered and more intensity is required to reach it. For example,
when $\Gamma \tau = 10$, $n_{p}$ has an asymptotic value of 1.84 and
is 78\% of this value when $I=I_{s}$. The saturation process
described here also leads to a type of power broadening of the
spectral line when the laser frequency is scanned. An increase in
laser intensity increases the fluorescence in the wings of the
resonance more than at the centre of the resonance, and so the line
is broadened. This is seen in the inset to
Fig.\,\ref{Fig:Saturation}, where the lineshape is plotted for the
various values of $\Gamma \tau$, in the case where $r=0.5$ and $I =
I_{s}/3$.

If the decay rate of the excited state happens to be small, it is
possible that the system will be in the opposite limit, $\Gamma
\tau \ll 1$. In that case, even if $R \gg \Gamma$, the mean number
of photons scattered by each molecule will be much smaller than 1,
and independent of $r$. We cannot then use
Eq.\,(\ref{Eq:photonsPerMolecule}) because the damping has no time to act. However,
 Eq.\,(\ref{Eq:single pulse}), which has no damping, can be integrated to give
\begin{equation}
n_{p} = \frac{\Gamma\tau}{2}\left(1-\frac{\sin(\Gamma \tau\sqrt{I/(2 I_s)})}{\Gamma\tau\sqrt{I/(2 I_s)}}\right)  \quad (\Gamma\tau \ll
1, R \gg \Gamma).
\end{equation}

\noindent The fluorescence saturates to the value $\Gamma \tau/2$
at high laser intensities.

Finally, it is interesting to consider the case where the
interaction time is long ($R\tau \gg 1$, $\Gamma \tau \gg 1$) and
$r$ is very close to 1 so that each molecule has the possibility of
scattering a large number of photons. Then,
Eq.\,(\ref{Eq:photonsPerMolecule}) reduces to the familiar atomic
physics result,

\begin{equation}
n_{p} = \frac{R\Gamma\tau}{2R + \Gamma} = \frac{\Gamma \tau}{2}
\frac{I/I_{s}}{1+I/I_{s}+4\delta^{2}/\Gamma^{2}}.
\end{equation}

\noindent The on-resonance fluorescence now saturates to
$\Gamma\tau/2$ when the condition $I \gg I_{s}$ is met.

Returning to the determination of the molecular flux, we write the
total number of photons detected per shot as

\begin{equation}
p = \frac{1}{L^{2}}\iint N(x,z) \epsilon(x,z) n_{p}(x,z)\,dx\,dz,
\label{Eq:PhotonsPerShot}
\end{equation}

\noindent where $N(x,z)$, $\epsilon(x,z)$ and $n_{p}(x,z)$ are the
number of molecules per steradian in the quantum state of interest,
the detection efficiency and the number of fluorescent photons per
molecule, all in the interval $dx\,dz$ around the point $(x,z)$ in
the plane of detection. To determine the molecular flux as
accurately as possible, the detection area should be defined by
placing a small aperture on the molecular beam axis, slightly
upstream of the detector, to block all but a small portion of the
molecular beam. Similarly, the probe laser should be collimated and
prepared with a top-hat intensity distribution in the $x$ direction.
Then $N$ and $\epsilon$ will be constant over the detection area.
The value of $n_{p}$ will be independent of $x$, but because of the
Doppler shift, it remains a sensitive function of $z$, particularly
if the degree of saturation is not high. Under these conditions,
Eq.\,(\ref{Eq:PhotonsPerShot}) reduces to

\begin{equation}
p = \frac{H N \epsilon}{L}\int n_{p}(\theta)\,d\theta,
\end{equation}

\noindent where $\theta=z/L$ and $H$ is the height (along $x$) of
the detection area. The integral is straightforward to calculate by
substituting the Doppler shift $\delta = 2\pi v_{0}\theta/\lambda$
into Eq.\,(\ref{Eq:ExcitationRate}) and then using
Eqs.\,(\ref{Eq:photonsPerMolecule}) and (\ref{Eq:Rpm}) (the probe
laser, of wavelength $\lambda$, is assumed to be on resonance). The
calculation requires some knowledge of $r$, $\Gamma$ and $I_{s}$,
but provided the detection is in the saturated regime, rough
estimates will suffice since the result becomes rather insensitive
to $\Gamma$ and $I_{s}$, and also to $r$ if $r$ is small, as is
often the case.

For the detection setup shown in Fig.\,\ref{Fig:SupersonicSetup},
the total detection efficiency is

\begin{equation}
\epsilon=(\Omega_{\rm l} /4\pi) \sum_i q_i (1+{\cal
R}(\lambda_i))T_{\rm l}(\lambda_i)^2 \chi(\lambda_i)\,.
\end{equation}

\noindent Here, $q_i$ is the fraction of fluorescent photons in
the emission line whose wavelength is $\lambda_i$, ${\cal R}$,
$T_{\rm l}$ and $\chi$ are the wavelength-dependent mirror
reflectivity, lens transmission, and photodetector quantum
efficiency and $\Omega_{\rm l}$ is the solid angle subtended by
the light-gathering lens. If windows and filters are present in
the setup, their transmissions need to be included too.

By measuring the value of $p$ and performing the above
calculations, the flux of molecules in the detected quantum state
can be determined. With careful measurements, an uncertainty below
50\% should be possible. For the cold YbF molecules produced in
our laboratory, the flux is measured to be $1.4 \times 10^{9}$
ground state molecules per steradian per shot, when the carrier
gas is argon \cite{Tarbutt(1)02}.

\subsection{Rotational temperature}

The rotational temperature of the molecules can be determined by
scanning the laser frequency and recording the rotational spectrum.
When the rotational temperature is $T_{r}$, the intensity of a
rotational line in the spectrum is proportional to $N(J)\epsilon(J)
n_{p}(J)$, where $N(J)=(2J+1)\exp[-B J(J+1)/k_{B}T_{r}]$ is the
relative number of molecules in the rotational state $J$, $B$ is the
rotational constant, and $\epsilon(J)$, $n_{p}(J)$ express the
$J$-dependence of the detection efficiency and the number of
scattered photons per molecule. Usually, the efficiency $\epsilon$
is almost independent of $J$. In many cases this is also true of
$n_{p}$, making it easy to extract the rotational temperature from
the relative line intensities once a few rotational lines have been
measured. A more accurate temperature determination must take into
account the variation of the matrix elements with $J$.  For example,
in $^{1}\Sigma - ^{1}\Sigma$ transitions, the $M_{J}$-averaged value
of $D^{2}$ in the R-lines is proportional to $(J+1)/(2J+1)$, which
decreases from 1 to $1/2$ as $J$ goes from 0 to $\infty$. For the
P-lines, by contrast, $D^{2}\propto J/(2J+1)$, which increases from
0 to $1/2$. Since this variation of the matrix elements generally
affects both the excitation rate $R$ and the branching ratio $r$, it
influences the value of $n_{p}(J)$ in both the saturated and
unsaturated regimes. In the saturated regime, where
$n_{p}\simeq1/(1-r)$, the variation of $n_p$ with $J$ is strongest
if $r$ happens to be close to 1.

In our lab, we have measured rotational temperatures of YbF, CaF and
LiH beams. For the first two, the rotational temperatures are
usually very close to the translational temperatures (typically in
the range 1-5\,K) \cite{Tarbutt(1)02}. For LiH we measure rotational
temperatures considerably higher than the translational temperature
\cite{Tokunaga(1)07}.

\subsection{Source noise}

The flux of molecules obtained from these sources is subject to
shot-to-shot fluctuations, as well as having a slow drift (mostly
downward). The time scale for the slow drift is typically $10^{4} -
10^{5}$ shots on a given spot of the target. We usually attach the
target to the rim of a large disk, typically 20~cm in diameter,
which we rotate incrementally as each target spot becomes exhausted.
In this way, the lifetime of a target is of order $10^{7}$ shots.
The ability to run the source continuously for long periods of time
is very important for precision measurements. A second essential
requirement is that the shot-to-shot fluctuations be small, since
they can contribute directly to the noise in the experiment. In
measuring the electron electric dipole moment using cold YbF
molecules, our detector records approximately 3000 photons per shot,
with a corresponding $\sqrt{N}$ photon shot-noise limit of 2\%.
Source fluctuations should ideally be kept below that level, but
this proves rather difficult to achieve; when optimized, the
short-time-scale fluctuations of our source are typically 2-3\%.

\section{Coherent manipulation of internal states\label{sec:CoherentManipulation}}
\subsection{Stark and Zeeman shifts of the hyperfine states}
In section \ref{subsec:molecular flux}, we discussed the excitation
of higher electronic states using laser light to drive optical
dipole transitions. The interaction was strongly damped by
spontaneous emission from the upper level; indeed, the molecules
were detected by means of the scattered photons. In this section, we
consider some ways to manipulate the hyperfine sublevels within the
ground state of the molecule. In contrast to optical transitions,
the coherences between these ground-state levels are not radiatively
damped because the spontaneous transition rates are very low for
transition frequencies in the sub-GHz range. The control of these
coherences provides a basis for exceedingly high precision
measurements of electric and magnetic fields and for measurements
such as that of the electron EDM.

\begin{figure}
\includegraphics{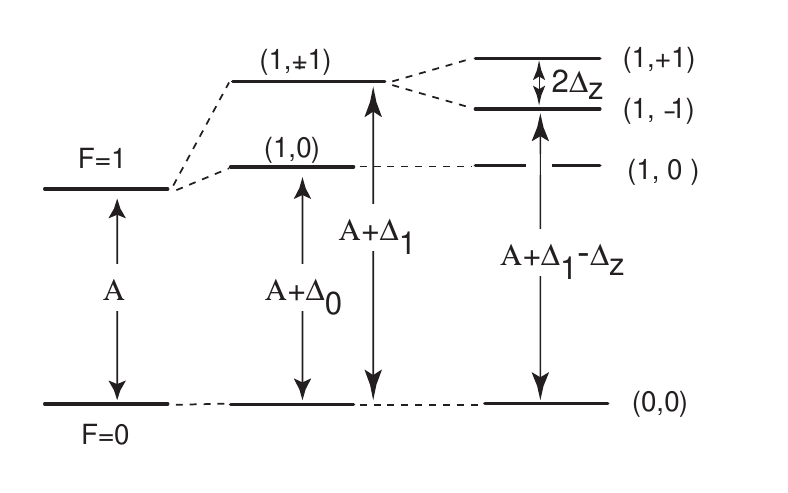}
\caption{Hyperfine levels. Left: field-free levels $F=0,1$. Centre:
electric field induced shift of levels and tensor Stark splitting of
the triplet $F=1$. Right: Zeeman splitting of the doublet
$F=1,M_F=\pm1$. \label{Fig:hfs}}\end{figure}

\begin{figure}[b]
\includegraphics{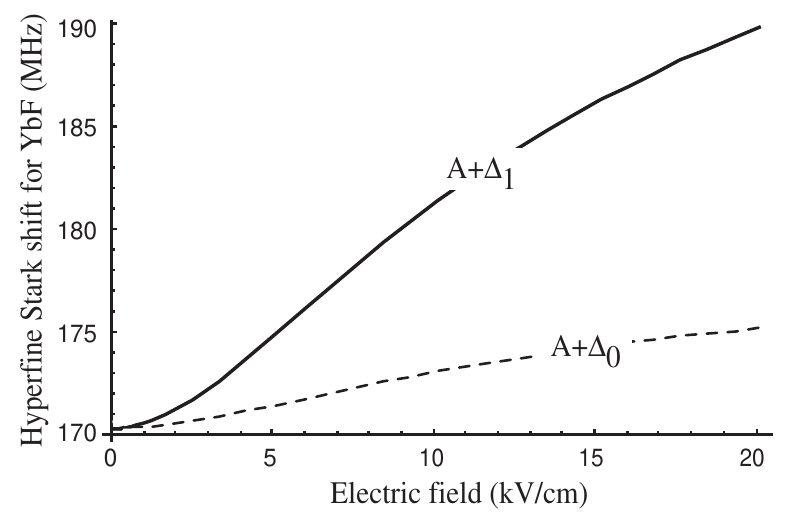}
\caption{Stark shift of the ground state hyperfine intervals in YbF.
\label{Fig:HyperfineStarkShift}}
\end{figure}

In order to be concrete in our discussion, let us take the simple
case illustrated in Fig.~\ref{Fig:hfs}, of two hyperfine levels,
$F=1$ and $F=0$, such as one finds in a diatomic X$^2\Sigma(N=0)$
molecule with nuclear spins of 0 and 1/2. Here the total angular
momentum $F$ is the sum of the electronic angular momentum $J=1/2$
and the nuclear spin $I=1/2$. Two examples of such molecules, which
have been studied in our laboratory, are $^{174}$YbF and CaF. In the
absence of any external fields, the two hyperfine levels are
separated by the hyperfine splitting $A$, as illustrated on the left
of Fig.~\ref{Fig:hfs}, and the three magnetic sublevels of $F=1$ are
degenerate. ($A=2\pi\times123$\,MHz in CaF and $2\pi\times170$\,MHz
in YbF). For a fuller discussion of hyperfine structure in such
molecules, see \cite{Sauer96}. When the molecule is subjected to an
electric field $E_z$, the main effect is the rigid rotor Stark shift
due to the electric dipole moment $\bm\mu_e$ along the internuclear
axis of the molecule, as we discuss more fully in Sec.\ref{Sec:AG}.
This shift is large in comparison with the hyperfine interaction;
for example the $N=0$ state of YbF shifts downwards by 20\,GHz at
20\,kV/cm. To a good approximation all four hyperfine levels shift
together, but there are some small differential shifts as well, due
mainly to the tensor part of the hyperfine interaction between $J$
and $I$, the electronic and nuclear angular momenta. This effect is
analysed in detail in Ref.\,\cite{Sauer96}. Relative to the $F=0$
level, the state $(F,M_F)=(1,0)$ shifts up by $\Delta_0$ whereas
states $(1,\pm1)$ shift up by $\Delta_1$, as illustrated in the
centre of Fig.~\ref{Fig:hfs}. These shifts are plotted in
Fig.\,\ref{Fig:HyperfineStarkShift} versus electric field for the
particular case of YbF. The lines are a calculation following the
theory detailed in \cite{Sauer96}, which has been confirmed by
experiment. One sees that these shifts of the hyperfine frequencies
within the $N=0$ manifold are typically a thousand times smaller
than the over-all shift of the manifold itself. The general
behaviour is a quadratic shift at low electric fields, where the
characteristic dipole interaction $-\bm\mu_e \cdot$\textbf{E} is
small compared with the rotational constant $B$, evolving to a more
nearly linear behaviour at higher fields and ultimately saturating
at the highest fields.

It is worth noting that the total Hamiltonian (including the
coupling to the external electric field) is invariant under time
reversal. Since the two states \{$(1,+1)$ in field $\textbf{E}$\}
and \{$(1,-1)$ in field $\textbf{E}$\} are time-reverses of each
other and since the Hamiltonian is invariant under time reversal, it
follows that the energy levels of $(1,\pm1)$ are exactly degenerate
at all electric fields. Thus, even though the molecule has an
electric dipole moment along its internuclear axis, and has an
induced electric dipole moment along the applied field direction
$z$, it does not have an electric dipole moment proportional to
$F_z$. This is a direct consequence of time reversal symmetry. By
contrast, if the electron were to have a permanent EDM
$\textbf{d}_e$ along its spin, this would lift the degeneracy
between the two levels - a direct consequence of the fact that such
an EDM violates time reversal symmetry. This spin-dependent Stark
shift is what our group measures in searching for an electron EDM
using YbF molecules \cite{Hud02a}.

A more mundane way to split the $(1,\pm1)$ levels, as illustrated on
the right of Fig.~\ref{Fig:hfs}, is with a static magnetic field
\textbf{B} through the Zeeman interaction $g_F
\mu_B\textbf{F}\cdot\textbf{B}$, where $\mu_B$ is the Bohr magneton,
and the g-factor expresses the ratio of magnetic moment to total
angular momentum. Provided this interaction is small compared with
the tensor Stark splitting $\Delta_1-\Delta_0$, the molecule only
responds to the component $B_z$ parallel (or antiparallel) to the
electric field. This splits the levels by $\pm\Delta_z=\pm g_F \mu_B
B_z/\hbar$. The additional contribution to this splitting resulting
from a perpendicular field component $B_\perp$ is of order
$\Delta_z[\mu_B B_\perp/(\Delta_1-\Delta_0)]^2$ \cite{Hud02a} and is
therefore negligible. No violation of time reversal symmetry is
implied here because the state \{$(1,-1)$ in field $\textbf{B}$\} is
not the time-reverse of \{$(1,+1)$ in field $\textbf{B}$\}: time
reversal also reverses the sign of \textbf{B}.  There is no shift
(to first order) of the states $(F,M_F)=(0,0)$ and $(1,0)$ since
they have $<F_z>=0$.

To summarise, the mean shift of the two levels $(1,\pm1)$ relative
to level $(0,0)$ is a measure of the electric field strength, whilst
the splitting between the two measures the parallel component of
magnetic field (and a possible very small contribution from the
electron EDM).  Although we have chosen to illustrate this with a
simple hyperfine system having $F=0$ and $F=1$, the same behaviour
applies more generally, namely that any two levels $F,\pm M_F$ are
equally shifted by the Stark interaction with electric field $E_z$,
whereas they are split apart by a magnetic field along $z$ (and by
an EDM). In the coherent manipulations we discuss below, the $(1,0)$
level plays no significant role and we therefore do not consider it
further. The remaining three levels can now be conveniently
abbreviated to $(0)$, $(+1)$ and $(-1)$, with energies 0,
$A+\Delta_1+\Delta_z$ and $A+\Delta_1-\Delta_z$ respectively. In the
next section, we will find it convenient to redefine the zero of
energy.

\subsection{Two-pulse interferometry of a 3-level
system}\label{subsec:Ramsey3Level}
 Let us write the amplitudes of
these three states as a column vector

\begin{equation}
a_z(t)=\left(\begin{array}{l}
a_0(t)\\
a_{+1}(t)\\
a_{-1}(t)\\
\end{array}\right)_z\, ,
\end{equation}
where the subscript $z$ indicates that the quantisation axis is
taken to be along the electric field. The free evolution of these
amplitudes from time $t_1$ to time $t_1+\tau$ in the presence of the
static electric and magnetic fields is given by the propagator

\begin{equation}
\label{Eq:pi0} \Pi_0(t_1,\tau)=\left(\begin{array}{lll}
e^{i\frac{\Omega}{2}\tau} & 0 & 0\\
0 & e^{-i(\frac{\Omega}{2}+\Delta_z)\tau} & 0\\
0 & 0 & e^{-i(\frac{\Omega}{2}-\Delta_z)\tau}\\
\end{array}\right)_z\,,
\end{equation}
such that $a_z(t_1+\tau)=\Pi_0(t_1,\tau)a_z(t_1)$. Here the
Stark-shifted hyperfine interval $A+\Delta_1$ has been replaced by
the symbol $\Omega$ and we have moved the zero of energy to the
centre of that interval in order to simplify the algebra that
follows.

Let us suppose that the molecules are prepared in state $(0)$, then
subjected to an rf magnetic field $\beta_x\cos(\omega t+\phi)$ along
$x$ in order to drive transitions to states $(+1)$ and $(-1)$. This
field excites the coherent superposition state
$(c)=\frac{1}{\sqrt{2}}[(+1)+(-1)]$ and does not couple at all to
the orthogonal superposition $(u)=\frac{1}{\sqrt{2}}[(+1)-(-1)]$.
(The converse is true for an rf field along $y$). This suggests a
new basis with quantisation along the $x$-axis, in which the new
state amplitudes are given by

\begin{align}\label{Eq:U}
a_x  = \left(\begin{array}{l}
a_0\\
a_{c}\\
a_{u}\\
\end{array}\right)_x & = U a_z\\
\nonumber & =\left(\begin{array}{lll}
1 & 0 & 0\\
0 & \frac{1}{\sqrt{2}} & \frac{1}{\sqrt{2}} \\
0 & \frac{1}{\sqrt{2}}  & -\frac{1}{\sqrt{2}} \\
\end{array}\right)\left(\begin{array}{l}
a_0\\
a_{+1}\\
a_{-1}\\
\end{array}\right)_z.
\end{align}

Note that the transformation $U$ is its own inverse: $U=U^{-1}$. In
the $x$-basis, only the states $(0$) and $(c)$ are coupled by the rf
field. With this reduction of the problem to a two-level problem, we
can write down how the amplitudes evolve in the $x$-basis under the
influence of the rf magnetic field applied from time $t_1$ to time
$t_1+\tau$. Following the standard derivation leading to Eq.~V.7 of
Ramsey's book\,\cite{Ramsey}, we find that
\begin{eqnarray}\label{Eq:rfPropagator}
\lefteqn{\Pi_{\rm{rf}}(t_1,\tau)=}\\
&\nonumber\left(\begin{array}{lll}
Z e^{i\frac{\omega}{2}\tau} & W e^{i\frac{\omega}{2}\tau}e^{i(\omega t_1 + \phi)} & 0\\
W e^{-i\frac{\omega}{2}\tau}e^{-i(\omega t_1 + \phi)} & Z^* e^{-i\frac{\omega}{2}\tau} & 0\\
0 & 0 & e^{-i\frac{\Omega}{2} \tau}\\
\end{array}\right)_x,
\end{eqnarray}
where
\begin{eqnarray}
\nonumber Z&=&i\cos\Theta\sin(\frac{a\tau}{2})+\cos(\frac{a\tau}{2})\\
\nonumber W&=&i\sin\Theta\sin(\frac{a\tau}{2})\\
\nonumber a&=&\sqrt{(\Omega-\omega)^2+4b^2}\\
\nonumber\cos\Theta&=&\frac{\Omega-\omega}{a}\\
\nonumber\sin\Theta&=&-\frac{2b}{a}\\
\nonumber b&=&\left<0\right|-\mu_x\beta_x\left|c\right>\,.\\
\nonumber\end{eqnarray} Although the phase of the field $\phi$ is
not important when a single pulse is applied, we keep it in this
formula because it becomes relevant when we consider double pulses.
As the state $(c)$ is excited, the effect of any static magnetic
field $B_z$ will be to rotate it into the third state $(u)$ at the
Larmor frequency $\Delta_z$. This has been ignored in deriving
Eq.~(\ref{Eq:rfPropagator}), under the assumption that the rf
excitation will be performed quickly in comparison with the Larmor
precession. Taken together, Eqs.~(\ref{Eq:pi0}), (\ref{Eq:U}) and
(\ref{Eq:rfPropagator}) provide us with all the tools we need to
investigate the evolution of this three-level system under any
sequence of short rf pulses.

When molecules in state $(0)$ are subjected to a single rf pulse,
the probability of excitation to state $(c)$ is given by
Eq.(\ref{Eq:rfPropagator}) as

\begin{eqnarray}
\label{Eq:single pulse}
\nonumber\lefteqn{P_{1\,pulse}(0\rightarrow c)=|W|^2}\\
&=\frac{4b^2}{(\Omega-\omega)^2+4b^2}\sin^2(\frac{\tau}{2}\sqrt{(\Omega-\omega)^2+4b^2})\,.
\end{eqnarray}

This is the usual magnetic resonance lineshape for transitions in a
2-level system without damping. At resonance the population
oscillates sinusoidally between the two states (this is known as
Rabi oscillation). A `$\pi$-pulse' is an on-resonance pulse with
$2b\tau=\pi$, which transfers all the population from state $(0)$ to
state $(c)$. In section \ref{subsec:OnePulseExpts} we will discuss
how this can be used in a molecular beam to map out the fields along
the beamline. An on-resonance `$\pi/2$-pulse' ($2b\tau=\pi/2$)
drives the transition only half way, creating an equal superposition
of states $(0)$ and $(c)$ with a definite relative phase. The
density matrix element describing this coherence at the end of the
pulse (at time $t_1+\tau$) is $(a_0 {a_c}^*
)_x=\frac{1}{2}i\exp\{i[\omega(t_1+\tau)+\phi]\}$, the phase of
which (apart from the fixed factor of $i$) is just the final phase
of the rf field.

In conventional Ramsey spectroscopy of a 2-level system
\cite{Ramsey}, two short $\pi/2$ pulses are applied in succession.
If the second pulse comes immediately after the first, the
transition is completed and all the population is excited. If
instead there is a delay time $T$ between the two pulses, which is
long compared to the pulse duration $\tau$, the transition
probability becomes sensitive to small differences between the rf
frequency and the molecular transition frequency. The internal
coherence evolving at the transition frequency accumulates a phase
between pulses of $\Omega T$, whereas the rf field evolves a phase
$\omega T$. When the difference between these two reaches $\pi$, the
second pulse reverses the effect of the first, returning all the
population to the initial state. More generally, the probability
that a molecule will end up in the excited state is
\begin{equation}
\label{Eq:Ramsey} P_{Ramsey}(0\rightarrow
1)=\frac{1}{2}\{1+\cos([\Omega-\omega] T-\delta\phi)\}\,,
\end{equation}
where we have allowed for the useful possibility of giving the
second pulse a phase $\phi+\delta\phi$ when the first has phase
$\phi$. These oscillations of the population, resulting from a beat
between the coherence and the driving field, are known as Ramsey
fringes. They are important because a long waiting time T allows a
small difference $\Omega-\omega$ to be measured, giving rise to very
precise spectroscopy.

We have generalised this idea to our system of three hyperfine
levels. The result, derived from Eqs.~(\ref{Eq:pi0}), (\ref{Eq:U})
and (\ref{Eq:rfPropagator}) is

\begin{eqnarray}
\label{Eq:doublepulse}
\lefteqn{P_{2\times pulse}(0\rightarrow 1)=}\\
&\nonumber1-\cos^2(b\tau)_1\cos^2(b\tau)_2-\cos^2(\Delta_z
T)\sin^2(b\tau)_1\sin^2(b\tau)_2\\
&\nonumber+\frac{1}{2}\cos([\Omega-\omega]T-\delta\phi)\cos(\Delta_z
T)\sin(2b\tau)_1\sin(2b\tau)_2\,.
\end{eqnarray}
Here we allow for the possibility that the values of $b\tau$ for the
two pulses, $(b\tau)_1$ and $(b\tau)_2$, are not equal.  In the
`Ramsey' case of two $\pi/2$-pulses, this simplifies to
\begin{eqnarray}
\label{Eq:2TimesPi/2}
\lefteqn{P_{2\times\frac{\pi}{2}\mbox{-\scriptsize
\it{pulse}}}(0\rightarrow 1)=}\\
\nonumber &\frac{1}{4}\left(3-\cos^2(\Delta_z
T)+2\cos([\Omega-\omega]T-\delta\phi)\cos(\Delta_z T)\right)\,.
\end{eqnarray}
If we set $\Delta_z$ to zero,  Eq.(\ref{Eq:2TimesPi/2}) reduces to
the standard two-level Ramsey result of Eq.(\ref{Eq:Ramsey}) because
the third state $(u)$ plays no role when the two levels $(+1)$ and
$(-1)$ are degenerate. When $\Delta_z\neq0$ it can be useful to pick
out the Ramsey interference by switching $\delta\phi$ between 0 and
$\pi$ and taking the difference, which is
$\cos([\Omega-\omega]T)\cos(\Delta_z T)$. As the frequency of the
oscillator is swept, the amplitude of these fringes provides
information about the Zeeman shift $\Delta_z$, while the phase of
the fringe pattern reveals the precise value of the splitting
$\Omega$ with a precision controlled by the choice of $T$. In
section \ref{subsec:TwoPulseExpts} we give an example of how this
can be used to look for small changes in a large electric field.

With three levels, it becomes possible to do interferometry using
$\pi$-pulses as well. Taking $2bt=\pi$, Eq.(\ref{Eq:doublepulse})
becomes
\begin{equation}
\label{Eq:Interferometer} P_{2\times\pi\mbox{-\scriptsize
\it{pulse}}}(0\rightarrow 1)=\sin^2(\Delta_z T)\,.
\end{equation}
This has a simple interpretation. Molecules are excited by the first
pulse to state $(c)$. This state subsequently evolves into
$\cos(\Delta_z T)(c)+i\sin(\Delta_z T)(u)$ because of the splitting
between states $(+1)$ and $(-1)$. The second pulse then drives the
state-$(c)$ part of the population back to $(0)$, leaving those in
state $(u)$ alone. As the magnetic field is scanned, this produces
fringes in the final state-$(0)$ population, with a  spacing that is
inversely proportional to $T$. These interference fringes can be
used for sensitive magnetometry and in searching for an electron
EDM.

\subsection{Experiments with single pulses\label{subsec:OnePulseExpts}}

\begin{figure}
\includegraphics{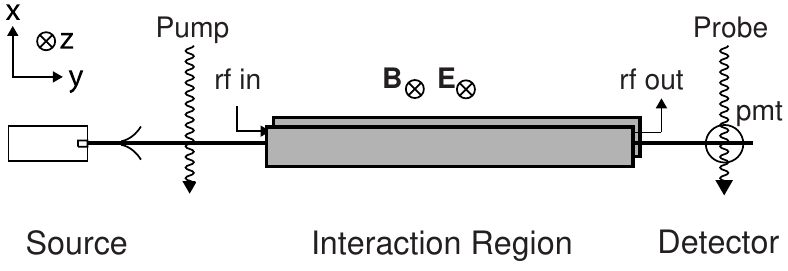}
\caption{Schematic diagram of the beam machine. Bunches of molecules
issue from the source (in the y-direction) and are skimmed before
being optically pumped into a single hyperfine state. The molecules
enter the magnetically shielded interaction region and fly through a
high-voltage capacitor where electric and magnetic fields can be
applied along $z$. This doubles as an rf transmission-line where the
rf magnetic field is along $x$. Finally, the molecules are detected
by laser induced fluorescence.
\label{Fig:InteractionRegion}}\end{figure}

Precision measurements usually require careful control and
monitoring of stray and applied fields, both electric and magnetic,
throughout the interaction region of the apparatus. The small
spatial and temporal extent of molecular beam pulses make it
possible to do so with high spatial resolution \cite{Hudson07}.
Fig.\,\ref{Fig:InteractionRegion} shows an apparatus to demonstrate
this using YbF molecules and pulsed rf fields. A pump laser and an
interaction region have been added in between the source and
detector already shown in Fig.\,\ref{Fig:SupersonicSetup}. Although
the beam is cold, both hyperfine levels are occupied because the
splitting $A=170$\,MHz is very much less than $kT$. The pump laser
excites the $A^2\Pi_{1/2}-X^2\Sigma^+$ Q(0) transition at 552~nm,
for which the Doppler width is (almost) eliminated by pointing the
laser beam perpendicular to the molecular beam. This makes the
excitation spectrum narrow enough ($\sim20$\,MHz) to excite just the
$F=1$ population so that it becomes selectively depleted. The
remaining $N=0$ molecules are virtually all in the $F=0$ state,
which serves as the initial state for subsequent manipulation by rf
pulses.

The interaction region starts 450~mm from the skimmer and is 790~mm
long. It is magnetically shielded to reduce the ambient field,
whilst current-carrying wires inside the shields can generate a
magnetic field if required. Within this region there is a pair of
electric field plates, 750~mm long and 70~mm wide, with a 12~mm
spacing, constant to better than 200~$\mu$m over the full length.
These are machined from cast aluminium, then gold coated  to improve
the uniformity of the surface potential (using a non-magnetic,
nickel-free process). The whole assembly is non-magnetic. With a
field across the gap of 15~kV/cm the leakage current is less than
1~nA.

\begin{figure}
\includegraphics{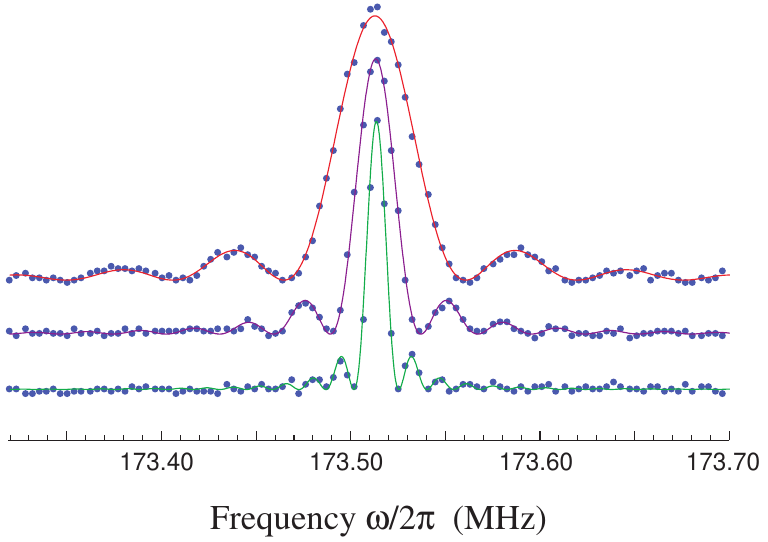}
\caption{Excitation spectra for the ground state $F=0 \rightarrow
F=1$ transition in YbF using $\pi$-pulses of three different pulse
lengths (upper, $\tau=18\,\mu$s; middle, $\tau=36\,\mu$s; lower,
$\tau=72\,\mu$s). A static electric field of 12.5\,kV/cm is applied.
Dots: experimental data. Lines: plots of Eq.\,\ref{Eq:single pulse}.
} \label{Fig:OnePulseLineshape}
\end{figure}

The same plate structure also serves as a 34~$\Omega$ transmission
line for the rf field, transporting it as a 170\,MHz TEM wave
travelling parallel or antiparallel to the beam direction. This is
described more fully in \cite{Hudson07}. As a bunch of molecules
travels along the interaction region, a hyperfine transition can be
induced at any desired position by pulsing the rf field on for a
short time. This repopulates the $F=1$ state and therefore produces
an increase in the fluorescence signal at the detector, as shown by
the rf frequency scans in Fig.\,\ref{Fig:OnePulseLineshape}.
Being a TEM wave, the rf magnetic field between the plates is
accurately perpendicular to the static electric field and we choose
to define its direction as the $x$-axis. This field therefore drives
the transition $(0)-(c)$ discussed in the last section.

Fig.~\ref{Fig:OnePulseLineshape} shows three excitation spectra,
obtained by applying $\pi$-pulses of three different durations (18,
36, and 72\,$\mu$s) to molecules near the middle of the interaction
region. Superimposed on the data points are solid lines
corresponding to Eq.~(\ref{Eq:single pulse}), which does a good job
of describing the lineshapes, including the positions and relative
heights of the sidebands. When we fix the value of $2b\tau$ at some
value $\theta$ (equal to $\pi$ in this case), these lineshapes can
be re-writen as a universal function
$\sin^2(\frac{\theta}{2}\sqrt{1+x^2})/(1+x^2)$, where
$x=(\Omega-\omega)\tau/\theta$. Thus, the width of the line is
inversely related to the duration of the pulse, becoming wider as
the pulse is made shorter, as one would expect from the usual
Fourier relation between pulse duration and spectral width. For a
$\pi$-pulse, the full width at half maximum is
$\delta\omega_{\rm{FWHM}}=5.0/\tau$.

\begin{figure}
\includegraphics{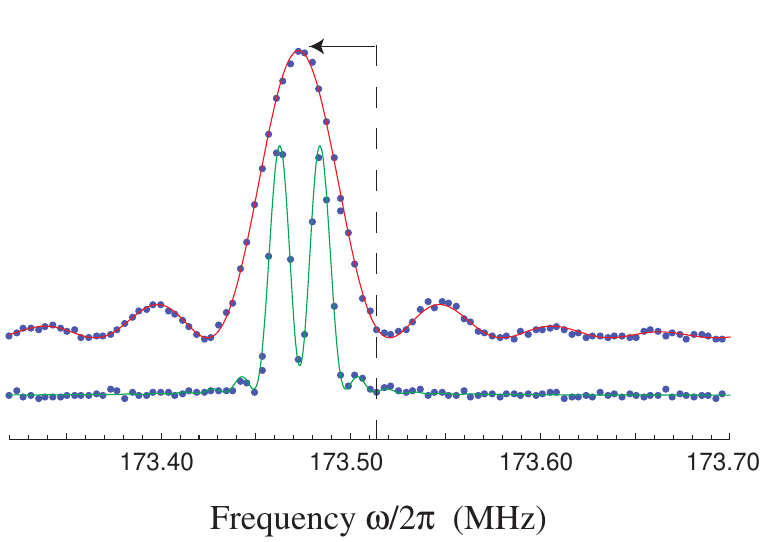}
\caption{Upper curve: the hyperfine transition moves to lower
frequency when the electric field is reduced. The dashed line marks
the line-centre of Fig.~\ref{Fig:OnePulseLineshape} at higher
electric field. Lower curve: The line splits when a magnetic field
is applied.}\label{Fig:OnePulseEandBshifts}
\end{figure}

The transition probabilities in Fig.~\ref{Fig:OnePulseLineshape}
peak at 173.513\,MHz, not at the field-free frequency of
170.254\,MHz, because these spectra were measured with a potential
difference of 15\,kV across the 12\,mm gap of the plates and are
therefore Stark-shifted by $\Delta_1$, as illustrated in
Fig.~\ref{Fig:hfs}. The magnitude of this shift is a measure of the
electric field strength at the place occupied by the molecules when
the rf pulse was applied. By varying the timing of the rf pulse, it
is possible to map out the electric field as a function of position
along the beamline \cite{Hudson07}. Note that the Doppler shift of
the rf transition is only a few hundred Hz and is therefore
insignificant at this level of accuracy. In
Fig.~\ref{Fig:OnePulseEandBshifts} we show how the resonance
frequency moves down (by 40.246\,kHz) when the applied potential
difference is reduced (by 203\,V). The spectral resolution afforded
by the $18\,\mu$s $\pi$-pulses is quite sufficient to see this shift
clearly. Fig.~\ref{Fig:OnePulseEandBshifts} also shows how the line
splits by $2\Delta_z$ (see Fig.~\ref{Fig:hfs}) when a DC magnetic
field of $0.8\,\mu$T is applied. In order to resolve the 22\,kHz
splitting fully, the linewidth was narrowed by increasing the pulse
duration to $72\mu$s. Over this time, a YbF molecule travelling at
590\,m/s covers 42\,mm, so it has been necessary to give up some
spatial resolution in order to achieve the higher spectral
resolution. The use of long rf pulses for high precision
spectroscopy is not ideal because the static field average that is
measured can be affected in a complicated way by variations in the
strength and polarisation of the rf field, either in space or in
time. As was first pointed out by Ramsey, it can be more
satisfactory to use a pair of rf pulses, and that is what we discuss
next.

\subsection{Experiments with double pulses\label{subsec:TwoPulseExpts}}
\begin{figure}
\includegraphics{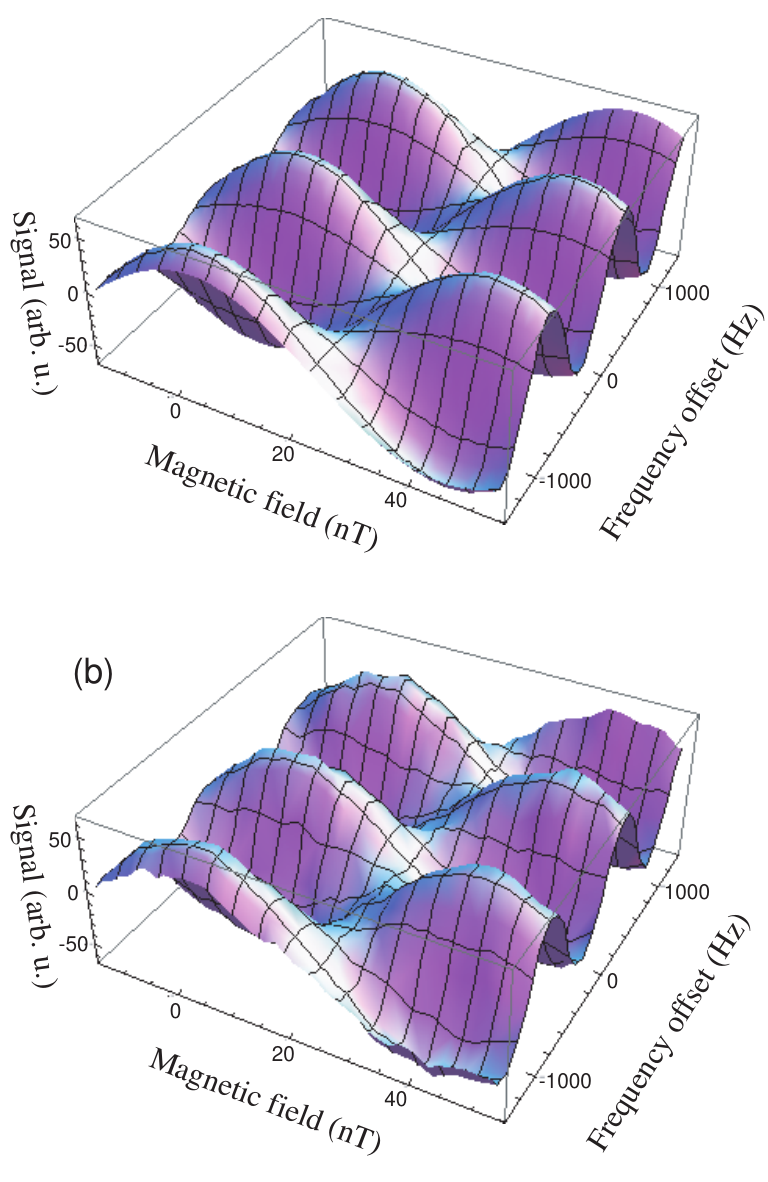}
\caption{Ramsey interference signals using two $\pi/2$-pulses
separated by a time $T=900\,\mu$s. The signals are plotted versus
the rf frequency $\omega$ and also as a function of magnetic field
in nT. (a) Theoretical signal, scaled to have the same amplitude as
the experimental result. (b) Measured Ramsey interference, in
excellent agreement with the theory. }\label{Fig:RamseyEB}
\end{figure}
In Eq.\,(\ref{Eq:2TimesPi/2}), we found an expression for the
lineshape using two short $\pi/2$-pulses separated by a time $T$. In
order to pick out the Ramsey interference, we suggested introducing
a phase shift $\phi$ between the first and second pulse and taking
the difference between $\phi=\pi$ and $\phi=0$. This is expected to
give fringes of the form $\cos(\Delta_z T)\cos([\Omega-\omega]T)$,
which we have plotted in Fig.~\ref{Fig:RamseyEB}(a) versus rf
frequency and magnetic field for the case when $T=900\,\mu$s. In
Fig.~\ref{Fig:RamseyEB}(b) we display the result of an experiment to
test this formula, in which we scanned the rf frequency $\omega$
many times through a region close to $\omega=\Omega$, stepping the
applied magnetic field $B_z$ in order to vary the Zeeman splitting
$2\Delta_z$. In this experiment, the molecules perform exactly as
predicted by Eq.\,(\ref{Eq:2TimesPi/2}).

\begin{figure*}
\includegraphics{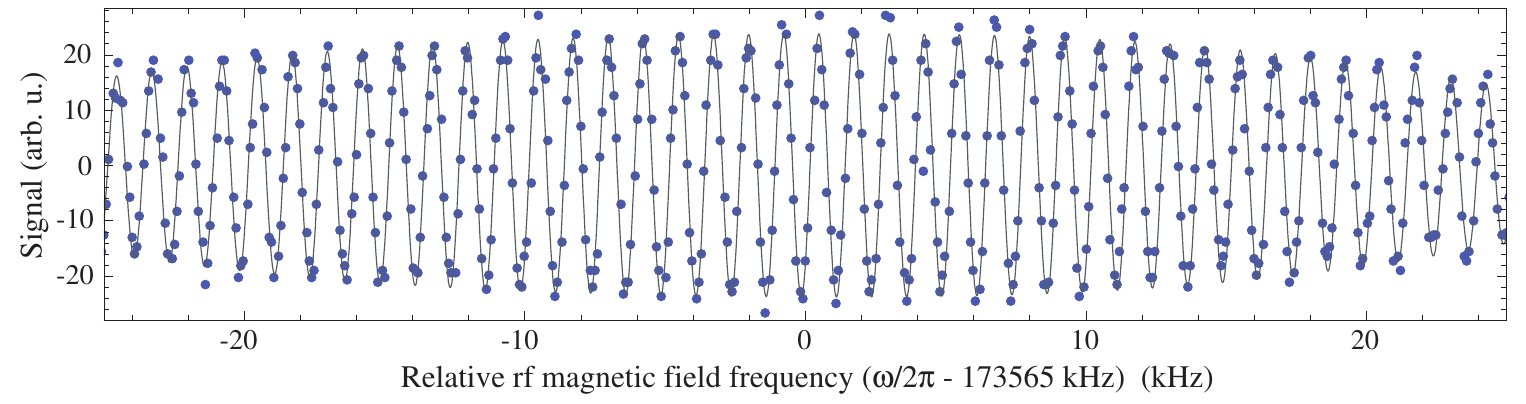}
\caption{The central 40 fringes of a Ramsey pattern obtained using
two $\pi/2$-pulses of $18\mu$s duration, separated by $T=800\,\mu$s.
The magnetic field is set very close to zero. Dots: Experimental
points. Line: Theoretical fringes with amplitude adjusted to fit the
data. }\label{Fig:RamseyE}
\end{figure*}

In Fig.~\ref{Fig:RamseyE}, we scan a much wider frequency range,
covering the central 40 Ramsey fringes with $B_z$ fixed close to
zero. Here we begin to see a departure from
Eq.\,(\ref{Eq:2TimesPi/2}) in the amplitude of the fringe pattern,
which shows a clear decrease on either side of the centre. This
happens because the individual pulses are starting to become
appreciably detuned, and therefore have reduced amplitude. Indeed,
when the detuning reaches $\pm \sqrt{15}/(4\tau)=\pm54$\,kHz, the
$18\,\mu$s single pulse transition probabilities go to zero and the
Ramsey interference has no amplitude at all. For a 2-level system,
the envelope of the interference pattern is just the single-pulse
lineshape, as discussed by Ramsey~\cite{Ramsey}. However, the
spectrum of Fig.~\ref{Fig:RamseyE} involves the third level $(u)$,
which makes it sensitive to very small magnetic fields and
complicates the shape of the envelope. This detuning effect is not
included in Eq.\,(\ref{Eq:doublepulse}), which assumes that the
detuning is small compared with $1/\tau$. At present we do not have
an analytical formula for this case, so the varying amplitude of the
$-\cos([\Omega-\omega]T)$ curve drawn through our data in
Fig.~\ref{Fig:RamseyE} is just a smooth fit to the measured
amplitude of the oscillations.
\begin{figure}[b]
\includegraphics{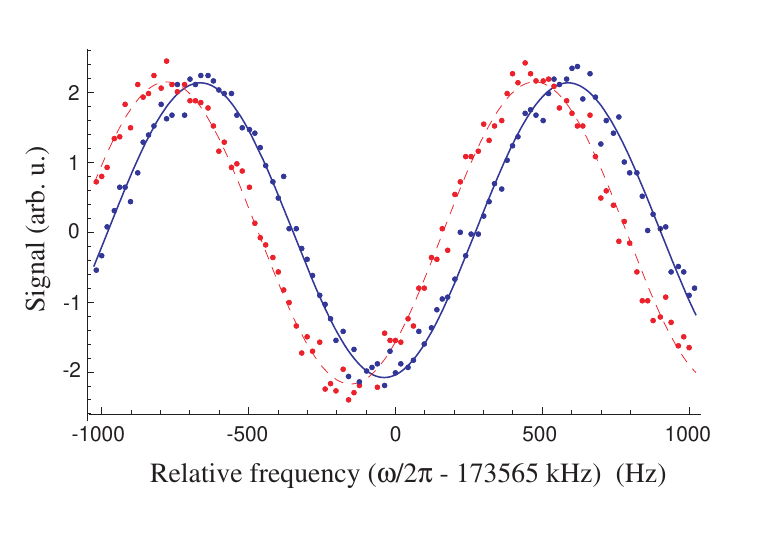}
\caption{The central fringe of Fig.~\ref{Fig:RamseyE}. Solid line:
fit to data taken with electric field as in Fig.~\ref{Fig:RamseyE}.
Dashed line: fit to data taken with high voltage leads reversed. The
evident Stark shift between the two indicates a change in the
magnitude of the electric field }\label{Fig:RamseyEBlowup}
\end{figure}

The curve in Fig.~\ref{Fig:RamseyEBlowup} shows a blow-up of the
central Ramsey fringe (solid line), together with the fringe
obtained when the high voltage leads were reversed (dashed line).
The small shift in the phase of the fringe pattern shows that the
12.5\,kV/cm electric field decreased in magnitude by
$455\pm11\,$mV/cm when we attempted to reverse it. This experiment
demonstrates  that Ramsey interferometry of the hyperfine states can
provide very sensitive monitoring of high electric fields in a
molecular beam apparatus.

\begin{figure}[b]
\includegraphics{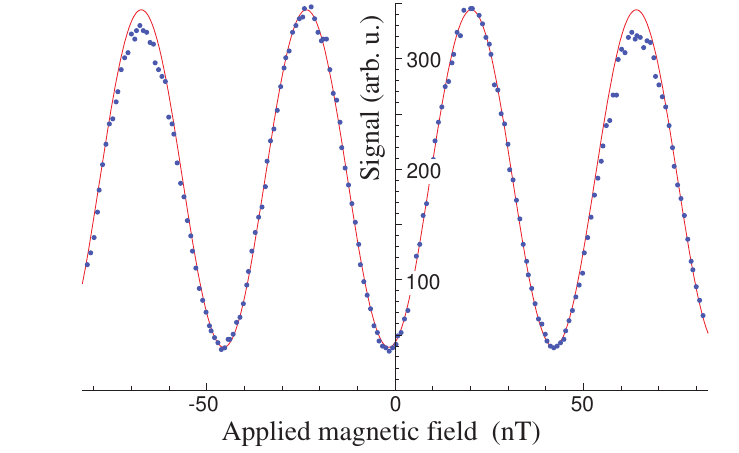}
\caption{Interferometer fringes obtained by scanning the applied
magnetic field, using a pair of $\pi$-pulses, separated in time by
$800\,\mu$s. Dots: experimental data. Line: a fit of the central two
fringes to the form $A\sin^2(\Delta_z
T)+C$.}\label{Fig:InterferometerFringes}
\end{figure}

We turn now to interferometry using pairs of $\pi$-pulses. Whereas
the $\pi/2$-pulses of Ramsey interferometry prepare and interrogate
an rf coherence between the states $(0)$ and $(c)$, the $\pi$ pulses
make and probe the superposition of the states $(+1)$ and $(-1)$,
which are almost degenerate. As expressed in
Eq.\,(\ref{Eq:Interferometer}), this superposition evolves at the
$(\pm 1)$ splitting frequency, leading to fringes of the form
$\sin^2(\Delta_z T)$, where $\hbar\Delta_z=g_F\mu_B B_z$.
Figure~\ref{Fig:InterferometerFringes} shows the interferometer
fringes measured in an electric field of 10\,kV/cm by scanning the
applied magnetic field over $160\,$nT, with a separation of
$T=800\,\mu$s between the two $\pi$-pulses. The solid line is a fit
over the two cental fringes to the lineshape $A\sin^2(\Delta_z
T)+C$, where $C$ represents the background due to unpumped $F=1$
molecules and to scattered light. Within the 2\% uncertainty of the
applied magnetic field calibration, the fringe spacing is found to
be 44\,nT, corresponding to a g-factor of $g_F=1$, as one would
expect for the $F=1$ hyperfine level of this $^2\Sigma$ state. The
total field $B_z$, which determines $\Delta_z$, is taken in our fit
to have an adjustable offset in addition to the applied field. This
is found to be $1.7\,$nT in Fig.~\ref{Fig:InterferometerFringes},
which is typical of the field that leaks through the magnetic
shielding from outside. While this simple function fits the central
two fringes well, the next fringe on either side clearly has less
amplitude. This is because the magnetic field detunes the two
$\pi$-pulse transitions from resonance.

Adjusting the magnetic field to reach a steep part of the fringe
pattern, the interferometer becomes sensitive to any small change in
the splitting $2\Delta_z$ between the levels $(\pm 1)$. This can be
used to monitor variations of the magnetic field in the apparatus.
In addition, if the electron has a dipole moment $d_e$, then the
interaction with the applied electric field $E$ also contributes to
$\Delta_z$, the additional amount being $\pm d_e \eta E/\hbar$. Here
$\eta$ is the enhancement factor discussed in
Sec.\ref{subsec:SymmetryTests}. If the electric field is reversed,
this shift changes sign, causing the fringe pattern to be modulated
from side to side as the electric field is flipped back and forth.
In this way, a measurement of the signal on the side of a fringe can
become very sensitive to the presence of a small electron EDM.

\section{Alternating gradient deceleration of polar molecules\label{Sec:AG}}
\subsection{Introduction} We have seen in section~\ref{sec:CoherentManipulation}
how spectral resolution can be improved by increasing the time $T$
available for coherent evolution. In a molecular beam with a
velocity of a few hundred m/s and an interaction region 1\,m long,
this time is a few ms, corresponding to minimum linewidths of
several hundred Hz.  For this reason, the prospect of decelerating
and trapping molecules offers significant improvements for some
precision measurements, provided the beam flux can remain high and
the inhomogeneous broadening due to trapping fields can be kept
under control. The basic idea of deceleration is to manipulate polar
molecules in an electric field gradient using the force due to the
Stark effect. After Stark deceleration was first demonstrated
\cite{Bethlem99}, the new technology was rapidly applied to make new
measurements. Using a beam of $^{15}$ND$_{3}$ molecules decelerated
to 52\,m/s, the energies of all 22 hyperfine levels of the
(J,K)=(1,1) state were measured with accuracies better than 100Hz
\cite{Vel04a}. Using Stark-decelerated OH radicals, greatly improved
measurements of the ground-state $\Lambda$-doublet microwave
transitions were made, thus contributing to the constraint on the
evolution of the fine-structure constant over cosmological time
\cite{Hud06a}. Once molecules could be trapped, it became possible
to measure directly the lifetimes of long-lived molecular states
\cite{Meerakker(1)05}, and even to measure the optical pumping of
molecules by room temperature blackbody radiation
\cite{Hoekstra(1)07}, which typically occurs on a timescale of many
seconds.

Many of the precision measurements discussed in this chapter make
use of heavy polar molecules. Stark deceleration of these heavy
species is considerably more challenging because (i) the kinetic
energy to be removed is proportional to the molecular mass and (ii)
the low-lying energy states are all high-field-seeking whereas the
Stark deceleration method works best for low-field seekers. The
first difficulty arises because the molecules formed in a supersonic
expansion acquire the speed of the carrier gas into which they are
seeded; thus their speed is independent of their mass. This
difficulty could be mitigated by using a low-temperature effusive
source such as the buffer-gas sources recently demonstrated
\cite{Maxwell(1)05, Patterson(1)07}.

\begin{figure}
\includegraphics{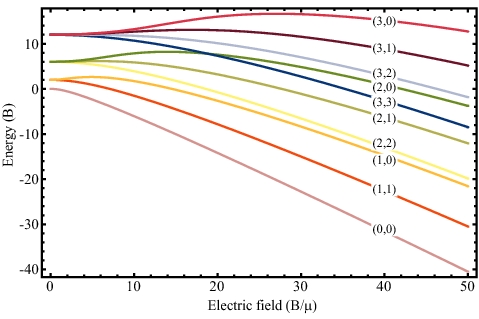}
\caption{The Stark shift of the low-lying energy levels of a rigid
rotating molecule. The electric field is expressed in units of
$B/\mu$ and the energy in units of $B$. States are labelled by the
quantum numbers $(J,M)$. \label{Fig:RotorStark}}\end{figure}

The second difficulty results from the closely packed rotational
energy levels of a heavy molecule which causes all the low-lying
states to be high-field seeking when the electric field is strong.
This problem is best illustrated by considering the Stark shift of a
rigid rotating molecule of reduced mass $m'$, bond-length $R$ and
dipole-moment $\mu$. The Hamiltonian is $H = B {\vec J}^{2} - \vec
\mu \cdot \vec E$, where $B=\hbar^{2}/(2 m' R^{2})$ is the
rotational constant, $\vec J$ is the angular momentum vector and
$\vec E$ is the applied electric field. Figure \ref{Fig:RotorStark}
shows the first 16 energy eigenvalues, in units of $B$, as a
function of applied electric field, in units of $B/\mu$. The field
mixes states having different values of $J$ but the same value of
$M$, the projection of the angular momentum onto the electric field
axis. Each energy level is labelled according to the quantum numbers
$(J,M)$ that the state evolves into when the electric field is
adiabatically reduced to zero. Note that the states $(J,M)$ and
$(J,-M)$ are degenerate for all electric fields, a consequence of
time-reversal-symmetry. The important point to note from
Fig.\,\ref{Fig:RotorStark} is that all the weak-field seeking states
have turning points, becoming strong-field-seekers at high field.
For example, the lowest-lying weak-field seeking state $(1,0)$ has
its turning point at an electric field of $4.9B/\mu$, at which point
the Stark shift is $0.64B$. Taking YbF as an example of a molecule
with a small rotational constant, the electric field at the turning
point is only 18\,kV/cm and the Stark shift only 0.15\,cm$^{-1}$.
This amount of energy, which can be removed from the molecule in a
single stage of deceleration, is to be compared with the
682\,cm$^{-1}$ of kinetic energy possessed by a YbF molecule formed
in a supersonic expansion at 290m/s \cite{Tarbutt(1)02}. Clearly a
very large number of stages would be needed to decelerate in this
way. By contrast, the strong-field-seeking ground-state of YbF has a
Stark shift of 10.7\,cm$^{-1}$ at a field of 200\,kV/cm, and so
Stark deceleration in this state seems feasible. For more complex
molecules, such as those of biological interest, the situation is
even more extremely weighted in favour of the high-field-seekers, as
discussed in more detail in \cite{Bethlem(1)06}.

Unlike weak-field-seekers, which are naturally focussed onto the
axis of the Stark decelerator, strong-field-seeking molecules cannot
be focussed using static fields. A dynamic focussing scheme needs to
be employed to prevent the molecules being pulled towards the
surfaces of the electrodes, where the field is strongest. The
alternating gradient focussing technique solves this problem of
transverse confinement. The molecules travel through a sequence of
electrostatic lenses, each of which focusses the molecules in one of
the two transverse directions and defocusses them in the other. The
focussing and defocussing planes alternate between one lens and the
next. For a subset of the molecules that enter the decelerator,
namely those that lie within the transverse phase-space acceptance
of the lens array, the net effect is to focus in both transverse
planes. Ideally, the focussing and defocussing forces are linear in
the off-axis displacements, and the trajectories of the accepted
molecules take them far from the axis inside the focussing lenses,
but close to the axis inside the defocussing lenses. Thus, the
overall focussing is a direct result of the motion of the molecules,
and can operate even when the defocussing power is stronger than the
focussing.

The first experiment to demonstrate alternating gradient
deceleration of polar molecules used an array of 12 lenses to
decelerate high-field-seeking metastable CO molecules from 275 to
260\,m/s \cite{Bethlem(1)02}. Ground-state YbF molecules were
decelerated from 287 to 277\,m/s using a similar machine
\cite{Tarbutt(1)04}.  The transverse focussing properties of the
alternating gradient have been demonstrated \cite{Bethlem(1)06} by
imaging metastable CO molecules exiting from the decelerator. Longer
machines, with more sophisticated electrode designs, should be
capable of decelerating heavy polar molecules to rest.

\subsection{A model alternating gradient decelerator}

\begin{figure}
\includegraphics{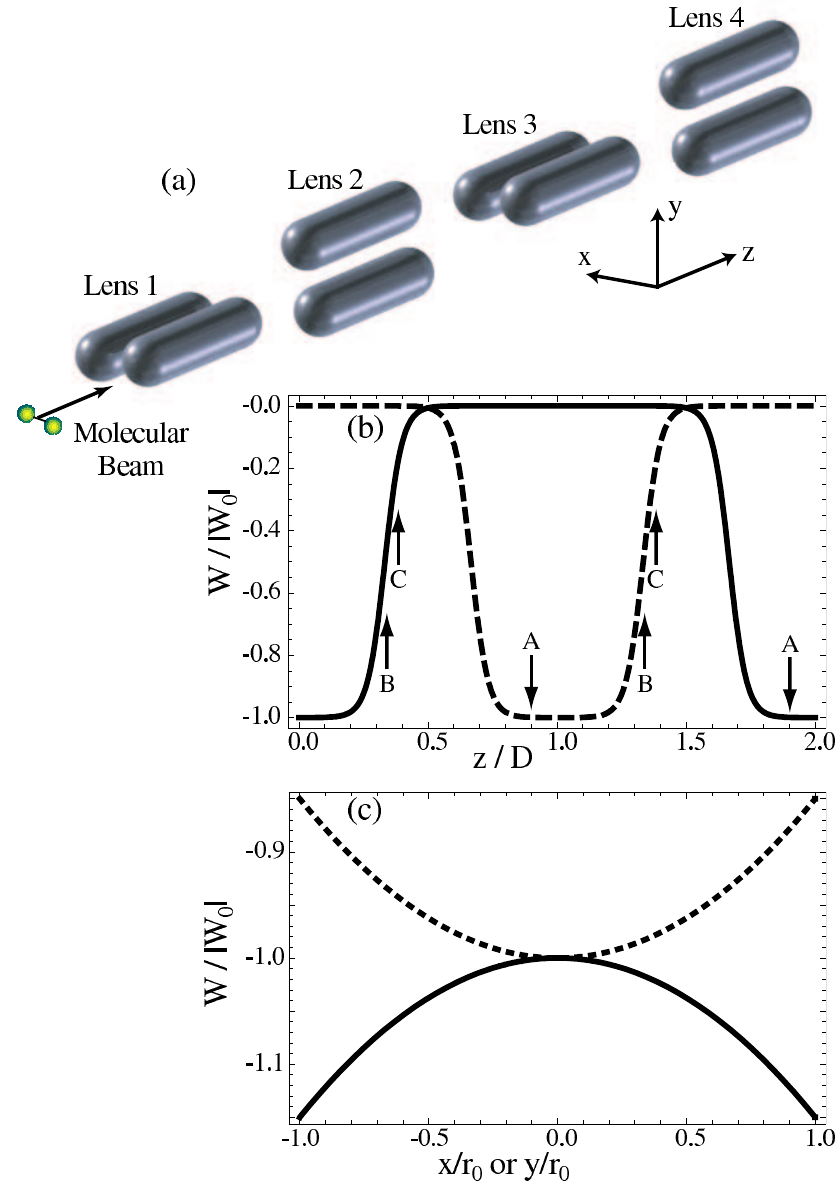}
\caption{(a) Schematic of a typical electrode structure in an
alternating gradient decelerator. (b) Potential in which the
molecules move, approximated using an analytical form (see text).
Solid and dashed lines represent respectively the switch states (i)
and (ii) of the decelerator. Positions A, B and C are referred to in
the text. (c) At the centre of the first lens, the potential along
$x$ is shown by the solid line and that along $y$ by the dashed
line. \label{Fig:ModelPotentials}}\end{figure}

We consider a decelerator consisting of a series of electrostatic
lenses whose focussing and defocussing planes alternate. A typical
electrode structure is shown in Fig.\,\ref{Fig:ModelPotentials}(a).
Here, each lens is formed by applying a large potential difference
between a pair of rods whose axes lie parallel to the beamline. The
forward velocity of the molecules only changes when they pass
through the fringe fields formed in the gap that separates one lens
from the next. In operation, the decelerator is switched between
three states: (i) odd lenses at high voltage, even lenses grounded,
(ii) even lenses at high voltage, odd lenses grounded and (iii) all
lenses grounded. We would like to work out the dynamics of molecules
in such a decelerator. We therefore first generate a map of the
electric field in the decelerator using an electrostatics solver,
and then construct the interaction potential, using the electric
field dependence of the Stark shift such as is plotted in
Fig.\,\ref{Fig:RotorStark}. We emphasise that this is not the
electrostatic potential, but rather the Stark potential in which the
molecules move. The electrode geometry shown in
Fig.\,\ref{Fig:ModelPotentials}(a) is one of many possible
geometries, some of which have been discussed in detail in reference
\cite{Bethlem(1)06}. The same reference discusses the Stark shift
and shows that it varies linearly with the electric field magnitude
for most heavy molecules in the electric fields of a typical
decelerator. The potential in which the molecules move is similar in
form for all the electrode geometries and molecular states
considered, though they vary in their details. Here, we do not
specialize to a particular geometry or molecular state, but instead
elucidate the dynamics using an analytical form for the potential
that approximates the true potential produced in most cases.

We take the potential in switch state (i) to be

\begin{equation}
W(x,y,z)=W_{0}(1 + b
((x/r_{0})^{2}-(y/r_{0})^{2}))f(z),\label{Eq:StarkPotential}
\end{equation}

\noindent where

\begin{equation}
f(z) = \frac{\tanh(z'/d + L/2d) - \tanh(z'/d -
L/2d)}{2\tanh(L/2d)}\label{Eq:f}
\end{equation}

\noindent provides a good phenomenological approximation to the
actual $z$-dependence. In Eq.~(\ref{Eq:StarkPotential}), $W_{0}$ is
the electric field at the origin, $r_{0}$ is a measure of the
transverse aperture of the decelerator, and $b$ measures the
transverse curvature of the potential. In Eq.~(\ref{Eq:f}), $z' =
{\rm mod}(z-D,2D) - D$, where ${\rm mod}(m,n)$ is the remainder on
division of $m$ by $n$, $D$ is the lens-to-lens spacing, $L$
measures the length of a lens, and $d$ measures how rapidly the
potential changes at the exit of the lens. The potential in switch
state (ii) is simply $W(y,x,z-D)$.

Figure \ref{Fig:ModelPotentials}(b,c) shows the potential for the
case where $L=2/3D$, $d=1/15D$ and $b = 0.15$. We shall use
these parameters throughout. Part (b) is a plot of the potential
along the beamline, $W(0,0,z)$, for both switch state (i) (solid
line) and switch state (ii) (dashed line). Part (c) of the figure
shows the transverse potentials in the centre of a lens,
$W(x,0,0)$ (solid line) and $W(0,y,0)$ (dotted line). In our ideal
lens, these are everywhere equal and opposite.

\subsection{Axial motion}

Consider molecules travelling down the axis of the decelerator. To
calculate when we should switch the potentials we introduce the
concept of a synchronous molecule, which enters the decelerator with
speed $u_0$. We design the switching sequence so that this molecule
is always at the same relative position in the periodic array every
time the field is turned on (e.g. position 'A' in Fig. 14(b)) and
every time the field is turned off (e.g. position 'B' in Fig.
14(b)). We refer to these fixed positions as $z_{\rm{on}}$ and
$z_{\rm{off}}$. Since this molecule is always climbing potential
hills, it decelerates as it moves along the beamline, and the time
intervals between successive switches must be chosen to increase in
correspondence. The required time sequence can be constructed using
a simple algorithm. Consider the energy conservation equation that
applies between the turn-on and turn-off points, $z_{{\rm on}}$ and
$z_{{\rm off}}$,

\begin{equation}
\tfrac{1}{2}M u_{n-1}^{2}+W(z_{\rm on}) = \tfrac{1}{2} M (dz/dt)^{2}
+ W(z).
\end{equation}
Here, $z$ is the position of the molecule at time $t$. Rearranging
and then integrating gives us the relationship between the $n^{\rm
th}$ turn-off time, $t_{{\rm off},n}$, and the $n^{\rm th}$ turn-on
time, $t_{{\rm on},n}$

\begin{subequations}
\begin{equation}
t_{{\rm off},n} = t_{{\rm on},n} + \int_{z_{{\rm on}}}^{z_{{\rm
off}}}\frac{dz}{\sqrt{u_{n-1}^2+2(W(z_{{\rm on}}) - W(z))/M}},
\end{equation}

\noindent where $u_{n}$ is the speed of the synchronous molecule
immediately after the $n^{th}$ turn-off. The $(n+1)^{th}$ turn-on is
now found using

\begin{equation}
u_{n} = \sqrt{u_{n-1}^{2}+2(W(z_{{\rm on}})-W(z_{{\rm off}}))/M},
\end{equation}
\begin{equation}
t_{{\rm on},n+1} = t_{{\rm off},n} + \frac{D - (z_{{\rm
off}}-z_{{\rm on}})}{u_{n}}.
\end{equation}
\end{subequations}

\begin{figure}
\includegraphics{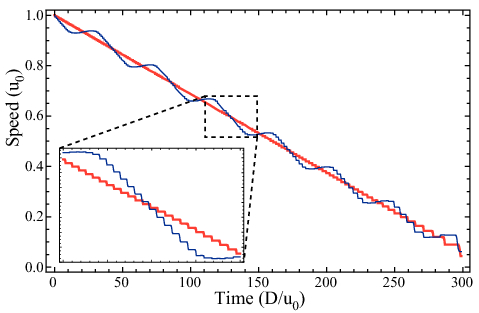}
\caption{The speed of a synchronous (thick red line) and a
non-synchronous molecule (thin blue line) as a function of time,
using the model potential and the turn-on and turn-off points A
and B indicated in Fig.\,\ref{Fig:ModelPotentials}(b). The inset
reveals the step-like structure of the deceleration.
\label{Fig:AxialSpeeds}}\end{figure}

Since we now have the potential seen by the molecules as a function
of both position and time, we can solve the equation of motion
numerically for any molecule, synchronous or not. Let us define $N$
to be the minimum number of stages required to stop the synchronous
molecule of mass $M$ and initial speed $u_{0}$, through the relation
$|W_{0}| = M u_{0}^{2}/(2N)$. Figure \ref{Fig:AxialSpeeds} shows the
result of such a calculation for the case where $L=2/3D$, $d=1/15D$
and $N=80$, and the turn-on and turn-off points are `A' and `B' in
Fig.\,\ref{Fig:ModelPotentials}(b). The thick red line shows the
speed of the synchronous molecule versus time, while the thin blue
line shows how the speed changes with time for a molecule that has
the same initial speed as the synchronous molecule but starts out
ahead by $D/15$. The deceleration of the synchronous molecule
appears to be uniform, though when magnified, as in the inset, is
actually seen to be a series of small steps reflecting the shape of
the potential. These steps can also be seen in the main figure when
the speed is low. The speed of the non-synchronous molecule
oscillates around that of the synchronous one and also consists of
many small steps when examined in detail. If we take snapshots of
the position and velocity of the molecules at each of the many
turn-off times, we would no longer be able to see the fine structure
of the motion. Between one snapshot and the next, the change in
kinetic energy of the synchronous molecule is $\Delta K = M u \Delta
u = M D u\Delta u/\Delta z_{s} = W(z_{{\rm on}}) - W(z_{{\rm
off}})$, where we have introduced the quantity $\Delta z_{s}$ to
represent the change in position of the synchronous molecule between
one snapshot and the next, and have used the fact that $\Delta z_{s}
= D$ always. Since we have discarded the information about the fine
structure of the motion, the deceleration of the synchronous
molecule appears to be constant, and we can convert to continuous
variables. The above equation then becomes $M D\,u\,du/dz_s = M
D\,d^{2}z_{s}/dt^{2} = W(z_{{\rm on}}) - W(z_{{\rm off}})$. Applying
a similar reasoning to any other molecule, described by the position
and velocity coordinates $z$ and $v$, we obtain $M D\,d^{2}z/dt^{2}
= W(z_{{\rm on}} + \tilde z) - W(z_{{\rm off}}+\tilde z)$, where
$\tilde z = z-z_{s}$, and we have made the additional approximation
that $v - u \ll u$ so that the distance moved by the general
molecule between switching times is also very close to $D$.
Subtracting the equation for the general molecule from that for the
synchronous molecule we obtain an equation of motion for the
relative coordinate, allowing us to define an effective force,
$F_{{\rm eff}}$:

\begin{align}
\frac{d^{2}\tilde z}{dt^{2}} &= \frac{W(z_{{\rm on}} + \tilde z) -
W(z_{{\rm on}}) - W(z_{{\rm off}}+\tilde z) + W(z_{{\rm off}})}{M
D}\nonumber \\
&= F_{{\rm eff}}/M.\label{Eq:EqOfRelAxMotion}
\end{align}

Introducing the relative velocity $\tilde v = v - u = d\tilde
z/dt$, the left hand side of the above equation can be written as
$\tilde v\,d\tilde v/d\tilde z$. Integrating, we then obtain

\begin{equation}
1/2 M \tilde v^{2} + (V(\tilde z) - V(0)) = E_{0},
\end{equation}

\noindent where $E_{0}$ is a constant and

\begin{equation}
V(\tilde z) = - \int F_{{\rm eff}}\,d\tilde z
\label{Eq:effectivePotential}
\end{equation}

\noindent is an effective potential for the relative motion
between non-synchronous and synchronous molecules.

\begin{figure}
\includegraphics{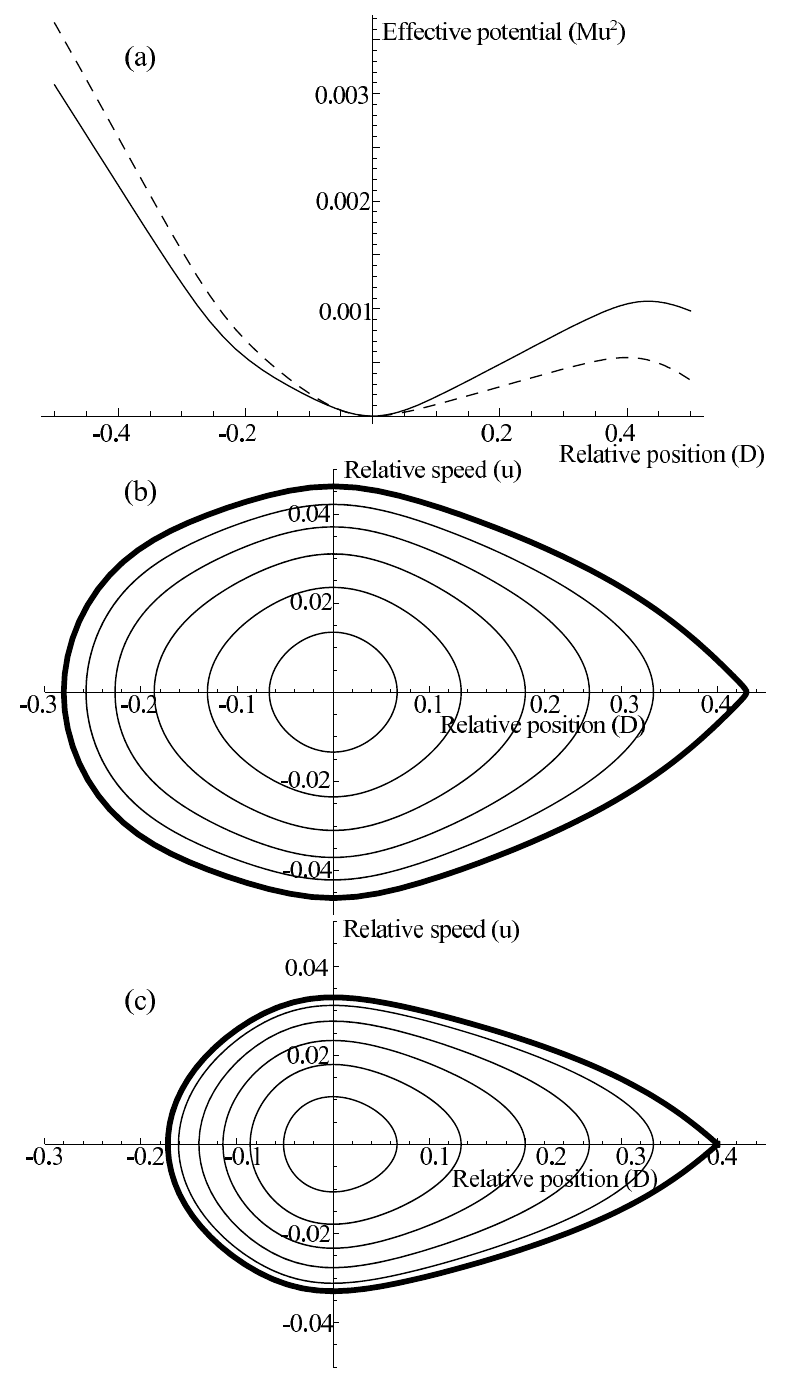}
\caption{(a) The effective potential,
Eq.\,(\ref{Eq:effectivePotential}), as a function of relative
position. The solid line corresponds to moderate deceleration
(turn off position `B'), while the shallower potential shown by
the dashed line is for strong deceleration (turn off position
`C'). The trajectories in phase-space calculated using these
effective potentials are shown in (b) and (c) for the deep and
shallow potentials respectively. The thick outer boundaries in
these plots are the
separatrices.\label{Fig:AxialPSPlots}}\end{figure}

Figure \ref{Fig:AxialPSPlots}(a) shows the effective potential
obtained from Eqs.\,(\ref{Eq:StarkPotential}), (\ref{Eq:f}) and
(\ref{Eq:effectivePotential}). The solid line corresponds to
$z_{{\rm on}}=-D/10, z_{{\rm off}} = D/3$, while the dashed line has
$z_{{\rm off}} = 11D/30$ corresponding to greater deceleration, but
a shallower effective potential. Since the effective potential is
confining, non-synchronous molecules with insufficient energy to
reach the top of the potential must oscillate about the synchronous
molecule. We can use the effective force to solve for the relative
motion of non-synchronous molecules. Parts (b) and (c) of
Fig.\,\ref{Fig:AxialPSPlots} show trajectories in phase-space
obtained in this way, for the solid and dashed effective potentials
shown in part (a). In each case, the thicker line separates bounded
and unbounded motion and is called the separatrix. All molecules
inside the separatrix will remain close to the synchronous molecule
throughout the deceleration process, and the area bounded by the
separatrix is the axial phase-space acceptance. Comparing parts (b)
and (c) of the figure, we see that, as expected from the shallower
potential, the acceptance is smaller when the deceleration is
greater. We note that the same phase-space plots can be generated
without making use of the effective potential, by numerically
integrating the complete equation of motion as was done in
generating Fig.\, \ref{Fig:AxialSpeeds}. The trajectories in phase
space then acquire the detailed structure shown in the inset of that
figure, but are otherwise found to be identical to those obtained
(much more rapidly) from the effective potential.

For small-amplitude oscillations about the synchronous molecule,
the motion is harmonic. Expanding the right hand side of
Eq.\,(\ref{Eq:EqOfRelAxMotion}) in a Taylor series about $\tilde z
= 0$, gives $d^{2}\tilde z/dt^{2} - (W'(z_{{\rm on}})-W'(z_{{\rm
off}}))\tilde z/(M D) = 0$ where $W'(a)=dW/dz$ evaluated at $z=a$.
The angular oscillation frequency for small-amplitude axial
oscillations is therefore

\begin{equation}
\omega_{z} = \sqrt{\frac{W'(z_{{\rm off}})-W'(z_{{\rm on}})}{M D}}.
\end{equation}

\noindent For our model potential, the solid line in
Fig.~\ref{Fig:AxialPSPlots}(a), the frequency can be conveniently
expressed as $\omega_{z}/2\pi = 0.309\,u_{0}/(\sqrt{N}D).$ For
example, if $D=30$\,mm,and $N=80$ when $u_{0}=300$\,m/s, we find
$\omega_{z}/2\pi=345$\,Hz for turn-off at position `B'.

\subsection{Transverse motion}

Our model potential is harmonic in the two transverse directions,
with a curvature that varies with $z$. As a molecule travels through
a lens, the curvature is very nearly constant until it reaches the
fringe-field of the lens where the curvature drops rapidly to zero.
To simplify the analysis, we make the approximation that the
curvature has the constant value, $W_{0}b/r_{0}^{2}$ over the
lens length $L$, is zero in the drift space of length $S=D-L$, and
changes abruptly between these values. As the decelerator structure
is periodic in $z$, it is natural to write the equation of motion
with independent variable $z$ rather than time. For a molecule with
forward speed $u$, the equation of motion is

\begin{equation}
d^{2}x/dz^{2} + \kappa^{2} Q(z) x = 0,
\label{Eq:TransverseEqOfMotion}
\end{equation}

\noindent where $Q(z) = 1$ inside a focussing lens, -1 in a
defocussing lens and 0 in a drift region, and the spatial frequency
is

\begin{equation}
\kappa = \sqrt{\frac{2W_{0}b}{M u^{2}r_{0}^{2}}} =
\sqrt{\frac{b}{N r_{0}^{2}}}. \label{Eq:kappa}
\end{equation}

\noindent The angular frequency of the transverse oscillation inside
a focussing lens is independent of the beam velocity $u$ and is
related to $\kappa$ by $\Omega = \kappa u$.

In moving through a region of length $l$, from an initial axial
position $z_{0}$, the transverse position and velocity coordinates
of a molecule change according to

\begin{equation}
\left( \begin{array}{cc} x/r_{0} \\ v_{x}/\Omega r_{0}
\end{array} \right)_{z_{0}+l} = M(z_{0}+l|z_{0})\left( \begin{array}{cc} x/r_{0} \\ v_{x}/\Omega r_{0}
\end{array} \right)_{z_{0}}.\label{Eq:transfer}
\end{equation}

\noindent Here, the dimensionless transfer matrix denoted by
$M(z_{0}+l|z_{0})$ takes the values $F(l)$ inside a focussing lens,
$D(l)$ inside a defocussing lens and $O(l)$ in a drift region, with

\begin{subequations}
\begin{align}
F(l) &= \left(
\begin{array}{cc} \cos(\kappa l)
& \sin(\kappa l) \\
-\sin(\kappa l) & \cos(\kappa l)
\end{array} \right),\\
D(l) &= \left(
\begin{array}{cc} \cosh(\kappa l)
& \sinh(\kappa l) \\
\sinh(\kappa l) & \cosh(\kappa l)
\end{array} \right),\\
O(l) &= \left(
\begin{array}{cc} 1
& \kappa l \\
0 & 1
\end{array} \right).
\end{align}\label{Eq:Matrices}
\end{subequations}

If, in moving from left to right, the molecule first travels through
a region described by the matrix $M_{1}$, followed by a region
described by $M_{2}$, the matrix in Eq.\,(\ref{Eq:transfer}) is
simply the product, $M=M_{2}\cdot M_{1}$. In this way, one complete
unit of our alternating gradient array is described by $M=F(L)\cdot
O(S)\cdot D(L)\cdot O(S)$, and a sequence of $N$ such units is
described (in a more compact but obvious notation) by $(FODO)^{N}$.
It can be shown that the molecular trajectories are stable if the
well known condition $-2<\,Tr(FODO)<2$ is satisfied (e.g.
\cite{Lee(1)99}).

We would like to know whether a molecule that enters the array
reaches the exit. For a long decelerator, the above stability
condition is a necessary but not sufficient one, since a molecule
can be on a stable trajectory that takes it so far from the axis
that it crashes into one of the electrodes. Rather than constructing
trajectories piecewise using Eqs.\,(\ref{Eq:Matrices}), we follow
the approach first used by Courant and Snyder in the context of the
alternating gradient synchrotron \cite{Courant(1)58}. We look for a
solution to Eq.\,(\ref{Eq:TransverseEqOfMotion}) of the general form

\begin{align}
x(z) &= \sqrt{\epsilon_{i} \beta(z)}\cos(\psi(z)+\delta_{i}) \nonumber \\
&= A_{1}\sqrt{\beta(z)}\cos \psi(z) + A_{2}\sqrt{\beta(z)}\sin
\psi(z) \label{Eq:trajEq}
\end{align}

\noindent where $\beta$ is a $z$-dependent amplitude function that
has the same periodicity as the AG array, $\psi$ is a
$z$-dependent phase, and $\epsilon_{i}$, $\delta_{i}$, $A_{1}$ and
$A_{2}$ are defined by the initial conditions. By substitution
into Eq.\,(\ref{Eq:TransverseEqOfMotion}) we find that
Eq.\,(\ref{Eq:trajEq}) is a valid solution provided

\begin{equation}
\psi(z) = \kappa \int_{0}^{z} \frac{1}{\beta(z')}\,dz'
\label{Eq:psi}
\end{equation}

\noindent and

\begin{equation}
-\frac{1}{4} \beta'^{2} + \frac{1}{2}\beta \beta'' + \kappa^{2}
Q(z) \beta^{2} = \kappa^{2}. \label{Eq:diffBeta}
\end{equation}

To find $\beta$, we need to make the connection to the piecewise
solution already given in Eqs.\,(\ref{Eq:Matrices}). From
Eq.\,(\ref{Eq:trajEq}) we have

\begin{equation}
x'(z) = \frac{A_{1}\kappa}{\sqrt{\beta}}(-\alpha \cos \psi - \sin
\psi) + \frac{A_{2}\kappa}{\sqrt{\beta}} (-\alpha \sin \psi + \cos
\psi) \label{Eq:xprime}
\end{equation}

\noindent where

\begin{equation}
\alpha = -\frac{1}{2\kappa} \beta'. \label{Eq:alpha}
\end{equation}

\begin{figure}
\includegraphics{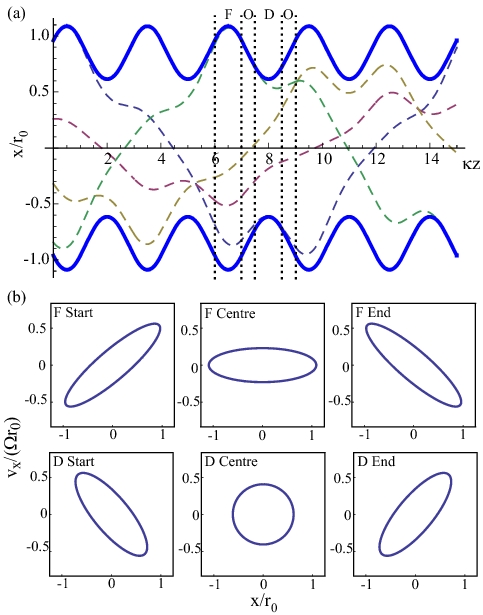}
\caption{(a) Dashed lines are typical trajectories through a
section of the alternating gradient array. Solid lines denote the
envelope that bounds the trajectories of all transmitted
molecules. The positions in the array of the focussing (F),
defocussing (D), and drift regions (O) are indicated by the dotted
lines. The figure shows that the beam envelope is largest at the
centre of the focussing lens, and smallest at the centre of the
defocussing lens. (b) Evolution of the phase-space ellipse through
a unit of the array. \label{Fig:Trajectories}}\end{figure}

Using Eqs.\,(\ref{Eq:trajEq}) and (\ref{Eq:xprime}) we find the
relationship between the coordinates $x,v_{x}/\Omega$ at position
$z$ and those at position $z+l_{{\rm cell}}$, $l_{{\rm cell}}=2D$
being the periodicity of the array. We make use of the periodicity
constraint on $\beta$, $\beta(z+l_{{\rm cell}}) = \beta(z)$ and
thus obtain:

\begin{equation}
M(z+l_{{\rm cell}}|z)= \left( \begin{array}{cc} \cos \Phi + \alpha
\sin\Phi
& \beta \sin \Phi \\
-\gamma \sin \Phi & \cos \Phi - \alpha \sin \Phi
\end{array} \right),
\label{Eq:CSM}
\end{equation}

\noindent where $\gamma = (1 + \alpha^{2})/\beta$ and $\Phi =
\psi(l_{{\rm cell}})$ is the phase advance per unit cell. Since the
integral in Eq.\,(\ref{Eq:psi}) is taken over a full period, $\Phi$
is independent of $z$. The matrix in Eq.\,(\ref{Eq:CSM}) is known as
the Courant-Snyder matrix. We can now equate this matrix to the
explicit form for the transfer matrix for one lattice unit, and so
obtain $\beta(z)$. For example, at a distance $z$ beyond the start
of a focussing lens, $M(z+l_{{\rm
cell}}|z)=F(z).O(S).D(L).O(S).F(L-z)$. We then obtain $\Phi$ using
the relation $\cos\Phi = \,Tr(M)$/2, and then find $\beta$ by
equating the upper right hand element of $M$ to that of the
Courant-Snyder matrix.

Figure \ref{Fig:Trajectories}(a) shows a few trajectories (dashed
lines) calculated using Eq.\,(\ref{Eq:trajEq}) with $\kappa L =
1$, $\kappa S = 0.5$, and arbitrarily chosen values for
$\epsilon_{i}$ and $\delta_{i}$. The motion is a product of two
periodic functions, one of wavelength $l_{{\rm cell}}$ and the
other of longer wavelength $2\pi l_{{\rm cell}}/\Phi$. It is often
the case that $\Phi \ll 2\pi$, in which case the modulation with
wavelength $l_{{\rm cell}}$ has a small amplitude and is called
the micromotion, while the much longer wavelength motion is called
the macromotion. For the case shown in the figure, $\Phi =
0.38\pi$ and the separation into a micromotion and a macromotion
is evident. If we consider a large collection of molecules, all
having different values of $\delta_{i}$ and $\epsilon_{i}$, the
only constraint being that every $|\epsilon_{i}| < \epsilon$, then
all the trajectories will be bounded by the envelope
$\pm\sqrt{\beta\epsilon}$. The bold lines in the figure show this
envelope. The figure shows that the beam size modulates with the
period of the array, reaching its maximum size in the centre of
every focussing lens, and its minimum size in the centre of every
defocussing lens. Since the confining and deconfining forces are
linear in the off-axis displacements, the defocussing lenses have
less effect on the beam than the focussing lenses, this being the
key to the stability of the alternating gradient array. As the
power of the lenses increases, so too does the depth of modulation
of the envelope until, at the stability boundary where $\Phi=\pi$,
the beam size becomes zero at the centre of the defocussing lens.

Using the first line of Eq.\,(\ref{Eq:trajEq}) and its derivative to
form the quantity $x^{2} + (\alpha x + \beta v_{x}/\Omega)^{2}$, we
find the invariant

\begin{equation}
\gamma x^2 + 2 \alpha x v_{x}/\Omega + \beta (v_{x}/\Omega)^2 =
\epsilon_{i}. \label{Eq:CSInvariant}
\end{equation}

\noindent This equation defines an ellipse in the phase-space
whose coordinates are $x$ and $v_{x}/\Omega$. The coordinates of
all molecules with the same value of $\epsilon_{i}$ but different
values of $\delta_{i}$ lie on this ellipse. Replacing
$\epsilon_{i}$ by $\epsilon$ gives us the ellipse that bounds the
entire set of molecules in the collection discussed above. The
shape of this ellipse changes periodically with $z$, but its area
is a constant, $\pi\epsilon$. Figure \ref{Fig:Trajectories}(b)
shows the phase-space ellipse at various positions in the array.
The beam is diverging as it enters the focussing lens. Inside this
lens the ellipse rotates, reaching its maximal spatial extent at
the lens centre where its principal axes are parallel to the
coordinate axes. The ellipse continues to rotate so that it is
converging at the exit of the focussing lens, and still converging
as it enters the defocussing lens. At the centre of the
defocussing lens, the spatial extent is minimized and the
principal axes of the ellipse are again along the coordinate axes.
The beam then starts to diverge again.

\begin{figure}
\includegraphics{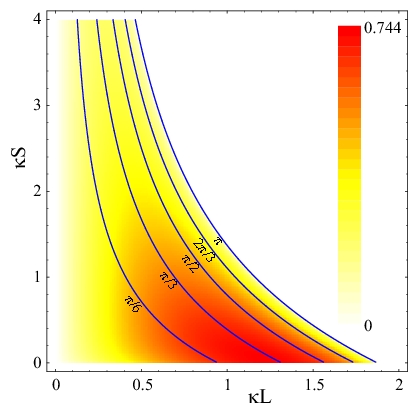}
\caption{Phase-space acceptance in one transverse direction, as a
function of $\kappa L$ and $\kappa S$. The acceptance is shown in
units of $\Omega r_{0}^{2}$.
\label{Fig:TransversePSAcceptance}}\end{figure}

Molecules will only be transmitted if their trajectories do not take
them outside the natural boundaries formed by the electrodes. The
characteristic size-scale in the transverse direction is $r_{0}$ and
so to calculate the transverse phase space acceptance, we assume
that the electrodes impose an aperture of size $2r_{0}$ in each of
the transverse directions. The beam characterized by $\epsilon$ will
be transmitted if $\sqrt{\beta \epsilon} < r_{0}$ everywhere. In
particular, the condition must be satisfied at the point where
$\beta$ has its maximum value, $\beta_{{\rm max}}$, which we already
know is at the centre of every focussing lens. The phase space
acceptance in $(x,v_{x})$ space is thus found to be $\pi
r_{0}^{2}\Omega/\beta_{\rm {max}}$. Figure
\ref{Fig:TransversePSAcceptance} is a density plot of the
phase-space acceptance in either transverse direction as a function
of the two dimensionless parameters that define the array, $\kappa
L$ and $\kappa S$. The region of highest acceptance is found near
$\kappa L \sim 1, S \ll L$, and the maximum value is
$0.744r_{0}^{2}\Omega$ obtained at $\kappa L = 1.254, \kappa S=0$.

The requirement that $\kappa L \sim 1$ for high acceptance
constrains the aspect ratio of the lenses, i.e. the ratio $L/r_{0}$.
Using Eq.\,(\ref{Eq:kappa}) and setting $\kappa L = 1$, we obtain
$L/r_{0} = \sqrt{N/b}$. Typical values for a decelerator with a
maximum field of 200\,kV/cm are $N=80$ and $b=0.15$, giving an
aspect ratio of $L/r_{0}=23.1$. Since $\kappa$ is inversely
proportional to $u$, it will increase as the molecules slow down. To
preserve the transverse acceptance, the lenses can be made
progressively shorter so as to maintain $\kappa L \sim 1$.
Alternatively, the alternating gradient array can have the structure
$(FO)^{n}(DO)^{n}$, with the value of $n$ decreasing down the
beamline.

\subsection{Beyond the ideal model}

In the previous section, we have used an idealized potential because
it allows us to understand the main aspects of the dynamics in a
straightforward way. This theoretical model misses some largely
undesirable effects that are present in reality. In the transverse
directions, nonlinear forces are necessarily present and these
reduce the transverse acceptance \cite{Kalnins(1)02, Bethlem(1)06, Tarbutt(1)08}.
The linear part of the force changes sign between one lens and the
next, leading to the dynamical stability discussed above, but the
leading order non-linear terms in the transverse force do not change
sign between the focussing and defocussing lenses and so tend to
upset the dynamical stability. Even when small compared to the
linear terms, the non-linear terms can significantly reduce the
transverse acceptance. Calculations for some typical electrode
geometries are presented in \cite{Bethlem(1)06}. In our idealized
model, the axial and transverse potentials are completely decoupled.
This cannot be achieved in any real decelerator because axial
gradients of the electric field, needed for deceleration, change the
dependence of the electric field on the transverse coordinates. In
particular, the fringe fields at the ends of the lenses tend to
increase the force constant in the defocussing direction relative to
the one in the focussing direction, leading to further beam loss
\cite{Bethlem(1)06}. A realistic simulation of the transmission of
molecules through an alternating gradient decelerator should use a
three-dimensional map of the electric field magnitude produced by
the electrode geometry of the machine, and the full electric-field
dependence of the Stark shift. Simulations of this kind show that
the true molecular trajectories in phase space are similar to those
calculated with our simplified model, though the acceptance volume
in phase space may be considerably smaller.

\section{Concluding remarks}In this
chapter we have summarised how precise measurements on molecules are
able to address important questions about the constancy of physical
laws and the structure of fundamental interactions. This is a
relatively new direction, which has emerged from a growing ability
to prepare and manipulate molecules in pure quantum states. We have
discussed the preparation of pulsed supersonic molecular beams and
have explored how coherent control of the hyperfine levels can
provide exquisite sensitivity to electric and magnetic fields. These
same methods also provide an opportunity to search within molecules
for more interesting effects, such as symmetry violations or
variation of fundamental constants, which can be related to new
physics on high energy scales. Deceleration and trapping are
important for improving these experiments because they can increase
the time available for coherent interaction with molecules from
milli-seconds to seconds. Although we have focussed on decelerating
heavy polar molecules using strong electric field seekers, the Stark
deceleration method can also be applied to weak-field-seeking polar
molecules and to Rydberg states of atoms and molecules.
Alternatively, it is possible to decelerate and trap molecules using
optical dipole forces. As methods for cooling, deceleration and
trapping molecules reach ever lower temperatures and higher
densities, and as they are extended to heavier and to more diverse species,
molecules will surely play an increasingly important role in testing
and understanding fundamental physics.


\begin{thebibliography}{99}
\bibitem{PulsedValve} To produce short pulses, we apply voltage pulses of duration 150-300\,$\mu$s and of amplitude 200-350\,V to a commercial solenoid valve (General Valve, Series 99, 50\,$\Omega$).
\end{thebibliography}
\end{document}